\documentclass[prd,twocolumn,preprintnumbers,nofootinbib,showpacs]{revtex4-1}
\usepackage[dvips]{color,graphicx}
\usepackage{bm}
\usepackage{amsmath,amssymb}

\newcommand{\MeV}{\;\text{MeV}}
\newcommand{\GeV}{\;\text{GeV}}

\begin{document}

\title{Chiral magnetic effect in the PNJL model}

\author{Kenji Fukushima}\email{fuku@yukawa.kyoto-u.ac.jp}
\author{Marco Ruggieri}\email{ruggieri@yukawa.kyoto-u.ac.jp}
\affiliation{Yukawa Institute for Theoretical Physics,
 Kyoto University, Kyoto 606-8502, Japan}

\author{Raoul Gatto}\email{raoul.gatto@unige.ch}
\affiliation{Departement de Physique Theorique,
 Universite de Geneve, CH-1211 Geneve 4, Switzerland}

\begin{abstract}
 We study the two-flavor Nambu--Jona-Lasinio model with the Polyakov
 loop (PNJL model) in the presence of a strong magnetic field and a
 chiral chemical potential $\mu_5$ which mimics the effect of
 imbalanced chirality due to QCD instanton and/or sphaleron
 transitions.  Firstly we focus on the properties of chiral symmetry
 breaking and deconfinement crossover under the strong magnetic field.
 Then we discuss the role of $\mu_5$ on the phase structure.  Finally
 the chirality charge, electric current, and their susceptibility,
 which  are relevant to the Chiral Magnetic Effect, are computed in
 the model.
\end{abstract}
\preprint{YITP-10-3}
\maketitle

\section{Introduction}

Quantum Chromodynamics (QCD) is widely believed to be the theory of
the strong interactions.  Investigations on its rich vacuum structure
and how the QCD vacuum can be modified in extreme environment are among the
major theoretical challenges in modern physics.  It is in particular
an interesting topic to study how non-perturbative features of QCD are
affected by thermal excitations at high temperature $T$ and/or by
baryon-rich constituents at large baryon (quark) chemical potential
$\mu_q$.  Such research on hot and dense QCD is important not only
from the theoretical point of view but also for numerous applications
to the physics problems of the Quark-Gluon Plasma (copiously produced
in relativistic heavy-ion collisions), ultra-dense and cold
nuclear/quark matter as could exist in the interior of compact stellar
objects, and so on.

The most intriguing non-perturbative aspects of the QCD vacuum at low
energy are color confinement and spontaneous breakdown of chiral
symmetry.  In recent years our knowledge on (some parts of) the QCD
phase diagram has increased noticeably because of significant
developments of the lattice QCD simulations
(see~\cite{deForcrand:2006pv,Aoki:2006br,Bazavov:2009zn,Cheng:2009be}
for several examples and see also references therein).  At zero
$\mu_q$, except for some reports~\cite{Aoki:2006br}, the numerical
simulations have almost established that two QCD phase transitions
(crossovers) take place simultaneously at nearly the same temperature;
one for quark deconfinement and another for restoration of chiral
symmetry (the latter being always broken because of finite bare
quark masses, strictly speaking).  It is still however under debate whether
two crossovers should occur at exactly the same temperature, however.

Once a finite $\mu_q$ is turned on, the Monte-Carlo simulation in
three-color QCD on the lattice cannot be performed straightforwardly
because of the (in)famous sign problem~\cite{Muroya:2003qs}.  To
overcome this problem several techniques have been developed such as
the multi-parameter reweighting method~\cite{Fodor:2001pe}, Taylor
expansion~\cite{Allton:2005gk}, density of state
method~\cite{Ambjorn:2002pz}, analytical continuation from the
imaginary chemical potential~\cite{D'Elia:2002gd}, the complex
Langevin dynamics~\cite{Aarts:2008wh}, etc.

In addition to hot and dense QCD with $T$ and $\mu_q$, the effect of a
strong magnetic field $\bm B$ on the QCD vacuum structure is also a
very interesting subject.  It would be of academic interest to
speculate modification of the vacuum structure of a non-Abelian
quantum field theory under strong external fields.  Besides, more
importantly, this kind of investigation has realistic relevance to
phenomenology in relativistic heavy-ion collisions in which a strong
magnetic field is produced in non-central
collisions~\cite{Kharzeev:2007jp,Skokov:2009qp}.  In particular, the
results obtained by the UrQMD model~\cite{Skokov:2009qp} show that
$eB$ created in non-central Au-Au collisions can be as large as
$eB \approx 2m_\pi^2$ (i.e.\ $B\sim 10^{18}$Gauss) for the top
collision energy at RHIC, namely
$\sqrt{s_{_{NN}}} = 200$~GeV.\ \ Moreover, an estimate with the energy
reachable at LHC, $\sqrt{s_{_{NN}}} \approx 4.5$~TeV, gives
$eB \approx 15 m_\pi^2$ for the Pb-Pb collision according to
Ref.~\cite{Skokov:2009qp}.  Hence, heavy-ion collisions provide us with
a most intriguing laboratory available on the Earth in order to study
the effect of extremely strong magnetic fields on the QCD vacuum.

Concerning the (electromagnetic) magnetic field effect on the QCD
vacuum structure, there have been many investigations and it has been
recognized that $\bm B$ plays a role as a catalyzer of dynamical
chiral symmetry
breaking~\cite{Klevansky:1989vi,Gusynin:1995nb,Fraga:2008qn,Klimenko:1990rh}.  The QCD
vacuum properties have been also studied by means of so-called
holographic QCD models~\cite{Zayakin:2008cy}.  The relation between
the dynamics of QCD in a strong magnetic field and non-commutative
field theories is investigated in Ref.~\cite{Gorbar:2005sx}.

A phenomenologically interesting consequence from the strong $\bm B$
in heavy-ion collisions is what is termed the Chiral Magnetic
Effect (CME)~\cite{Kharzeev:2007jp,Fukushima:2008xe}.  The underlying
physics of the CME is the axial anomaly and topological objects in
QCD.\ \ Analytical and numerical investigations have demonstrated that
the sphaleron transition occurs at a copious rate at high temperature
unlike instantons that are thermally
suppressed~\cite{Arnold:1987zg,Moore:1997im}.
Sphalerons are finite-energy solutions of the Minkowskian equations of
motion in the pure gauge sector and they appear not only in the
electroweak theory but also in QCD~\cite{McLerran:1990de}. They carry
a finite winding number $Q_W$ which is defined as
\begin{equation}
 Q_W = \frac{g^2}{32\pi^2}\int d^4 x \;\text{Tr}[F\tilde{F}] \;,
\end{equation}
where $F$ and $\tilde{F}$ denote respectively the field strength
tensor and its dual.  Sphalerons connect two distinct classical vacua
of the theory with different Chern-Simons numbers in Minkowskian
space-time.  It is possible through the coupling with fermions in the
theory to relate the change of chirality, $N_R - N_L$, to the winding
number by virtue of the Adler-Bell-Jackiw anomaly relation,
\begin{equation}
 (N_R - N_L)_{t=+\infty} - (N_R - N_L)_{t=-\infty} = -2 Q_W \;.
\label{eq:int1aaa}
\end{equation}
The r.h.s.\ of Eq.~\eqref{eq:int1aaa} is the integral over
space-time of $\partial_\mu j^\mu_5$, where $j^\mu_5$ represents the
anomalous flavor-singlet axial current.  The physical picture that
arises from Eq.~\eqref{eq:int1aaa} is that in the presence of
topological excitations such as instantons and sphalerons with a given
$Q_W$, and starting with a system of quarks with $N_R = N_L$, an
unbalance between left-handed and right-handed quarks is produced.
Such an unbalance can lead to observable effects to probe topological
$\mathcal{P}$- and $\mathcal{CP}$-odd excitations.  An experimental
observable sensitive to local $\mathcal{P}$- and $\mathcal{CP}$-odd
effects has been proposed in Ref.~\cite{Voloshin:2004vk}.  Recently
the STAR collaboration presented the conclusive observation of charge
azimuthal correlations~\cite{STAR} possibly resulting from the CME
with local $\mathcal{P}$- and $\mathcal{CP}$-violation.

The intuitive picture of the CME is as follows.  In a strong magnetic
field $\bm B$, quarks are polarized along the direction of $\bm B$.
Let us suppose that $\bm B$ is along the positive $z$ axis (that is
conventionally taken as the $y$ axis in the context of heavy-ion
collisions).  Neglecting quark masses, which is a good approximation
for $u$ and $d$ quarks in the high-$T$ chiral restored phase, the
chirality is an eigenvalue to label the quarks.  Then, right-handed
$u$ quarks should have both their spin and momentum parallel to
$\bm B$ and left-handed $u$ quarks should have their spin parallel to
$\bm B$ and momentum anti-parallel to $\bm B$.  Obviously the same
reasoning applies to $d$ quarks.  If $N_R = N_L$, then the current
that would originate from the motion of left-handed quarks is exactly
cancelled by that of right-handed quarks.  If $N_R \neq N_L$ which is
expected from the anomaly relation~\eqref{eq:int1aaa}, on the other
hand, a finite net current is produced.  Therefore, if quarks
experience a strong magnetic field in a domain where the topological
transition occurs, a net current is produced locally.

The CME has been investigated in the chiral effective
model~\cite{Nam:2009jb} as well as in the holographic QCD
model~\cite{Zayakin:2008cy}.  The chiral magnetic conductivity is
calculated without gluon interactions in Ref.~\cite{Kharzeev:2009pj}.
In~\cite{Fukushima:2009ft} the electric-current susceptibility under a
homogeneous magnetic field, which can be related to the fluctuation of
the electric-charge asymmetry measured by the STAR collaboration, has
been computed in the same way.  The first lattice-QCD study of the CME
has been performed by the ITEP lattice group~\cite{Buividovich:2009wi}
in the color-$\mathrm{SU}(2)$ quench approximation.  Moreover, the
Connecticut group~\cite{Abramczyk:2009gb} performed a lattice-QCD
study of the CME with $2+1$ dynamical quark flavors.

This article is devoted to the study of the two-flavor
Nambu--Jona-Lasinio model with the Polyakov loop coupling (PNJL
model) in a strong magnetic field.  The PNJL model has been introduced
in Refs.~\cite{Meisinger:1995ih,Fukushima:2003fw} to incorporate
deconfinement physics into the NJL model~\cite{revNJL}.  The main
addition to the NJL model is a background gluon field in the Euclidean
temporal direction.  The background field is related to the
expectation value of the traced Polyakov loop, $\Phi$, which is known
to be an order parameter for the deconfinement transition in a pure
gauge theory~\cite{Polyakovetal}.  There are many theoretical studies
related to different aspects of the PNJL model; see for
example Ref.~\cite{Ratti:2005jh}. 
See Refs.~\cite{Schaefer:2007pw} for a related study 
within the Polyakov-Quark-Meson model, and~\cite{Braun:2009gm}
for an investigation within QCD with imaginary chemical potential. 

We work in the chiral limit throughout the paper, in which the
definition of the chiral critical temperature has no ambiguity.
Firstly we focus on the effect of $\bm B$ on chiral symmetry
restoration at finite temperature.  As it will be clear soon, our
results support the role of the external $\bm B$ as a catalyzer of
dynamical symmetry breaking; the critical temperature increases with
increasing $\bm B$.  Naturally the (pseudo)critical temperature for
deconfinement crossover is less sensitive to the presence of $\bm B$
because there is no direct coupling between photons and gluons.
Hence, the PNJL model predicts that at large enough $\bm B$, a
substantial range of temperature will open at which quark matter is
deconfined but chiral symmetry is still dynamically broken.
See~\cite{Agasian:2008tb} for a related study. 

Also we shall discuss the effects of a finite chiral chemical potential
$\mu_5$ on the phase structure within the PNJL model.  This $\mu_5$
mimics the topologically induced changes in chirality charges
$N_5=N_R-N_L$ that are naturally expected by the QCD anomaly relation.
The relevant quantity in a microscopic picture is rather the total
chirality charge $N_5$ but for technical reasons it is easier to work
in the grand-canonical ensemble by treating $\mu_5$ (see
Ref.~\cite{Fukushima:2010vw} for an alternative description based on
the flux-tube picture), which is to be interpreted as the time
derivative of the $\theta$ angle of the strong interactions;
$\mu_5=\dot{\theta}/(2N_f)$.  Besides the phase diagram from the PNJL
model, we compute quantities that are relevant to the CME, namely, the
induced electric current density, its susceptibility, and the chiral
charge density $n_5$ together with its susceptibility.

This paper is organized as follows; in Sec.~\ref{sec:model} we give a
detailed description of the model we are using.  In
Secs.~\ref{sec:phase} and \ref{sec:cme} we present and discuss our
numerical results from the model.  Finally, in
Sec.~\ref{sec:conclusions} we draw our conclusions.

\section{Model with magnetic field and chiral chemical potential}
\label{sec:model}

In this section we analyze the interplay between the chiral phase
transition and the deconfinement crossover at large $\bm B$ using the PNJL
model.  Here we consider two-flavor quark matter in the chiral limit
since the chiral phase transition is a true phase transition only in
the chiral limit, and then and only then $T_c$ can be identified
unambiguously by vanishing chiral condensate.  The chiral limit in the
two-flavor sector is not far from the physical world in which the
current quark masses are a few MeV, almost negligible as
compared to the temperature.  Moreover, we are interested in studying
the situation in the presence of chirality charge density.  In the
grand-canonical ensemble we can introduce the chirality charge by
virtue of the associated chemical potential $\mu_5$ in the following
way.

The Lagrangian density of the model we consider is given by
\begin{align}
 \mathcal{L} &= \bar\psi\left(i\gamma_\mu D^\mu
  + \mu_5 \gamma^0\gamma^5 \right)\psi \notag\\
 &\quad + G\left[\left(\bar\psi\psi\right)^2
  + \left(\bar\psi i \gamma^5 \bm\tau\psi\right)^2\right]
  - \mathcal{U}[\Phi,\bar\Phi,T] \;,
\label{eq:four-fermi}
\end{align}
where the covariant derivative embeds the quark coupling to the
external magnetic field and to the background gluon field as well, as
we will see explicitly below.  We note that $\mu_5$ couples to the
chiral density operator
$\mathcal{N}_5=\bar{\psi}\gamma^0\gamma^5\psi
=\psi^\dagger_R \psi_R - \psi^\dagger_L \psi_L$, hence
$n_5 = \langle\mathcal{N}_5\rangle\neq0$ can develop when
$\mu_5\neq0$.  The mean-field Lagrangian is then given by
\begin{equation}
 \mathcal{L} = \bar\psi\left(i\gamma_\mu D^\mu - M
  + \mu_5 \gamma^0\gamma^5\right)\psi
  - \mathcal{U}[\Phi,\bar\Phi,T] \;,
\label{eq:Ika}
\end{equation}
where $M=-2\sigma$ with
$\sigma = G\langle\bar\psi\psi\rangle
 = G(\langle\bar u u\rangle + \langle\bar d d\rangle)$.

In Eq.~\eqref{eq:Ika} $\Phi$, $\bar\Phi$ correspond to the normalized
traced Polyakov loop and its Hermitean conjugate respectively,
$\Phi=(1/N_c)\text{Tr}L$ and $\bar\Phi=(1/N_c)\text{Tr}L^\dagger$,
with the Polyakov loop matrix,
\begin{equation}
 L = \mathcal{P}\exp\left(i\int_0^\beta \! A_4\, d\tau\right) \;,
\end{equation}
where $\beta=1/T$.

The potential term $\mathcal{U}[\Phi,\bar\Phi,T]$ in
Eq.~\eqref{eq:Ika} is built by hand in order to reproduce the pure
gluonic lattice data~\cite{Ratti:2005jh}.  Among several different
potential choices~\cite{Wambach:2009ee} we adopt the following
logarithmic form~\cite{Fukushima:2003fw,Ratti:2005jh},
\begin{equation}
 \begin{split}
 & \mathcal{U}[\Phi,\bar\Phi,T] = T^4\biggl\{-\frac{a(T)}{2}
  \bar\Phi \Phi \\
 &\qquad + b(T)\ln\bigl[ 1-6\bar\Phi\Phi + 4(\bar\Phi^3 + \Phi^3)
  -3(\bar\Phi\Phi)^2 \bigr] \biggr\} \;,
 \end{split}
\label{eq:Poly}
\end{equation}
with three model parameters (one of four is constrained by the
Stefan-Boltzmann limit),
\begin{equation}
 \begin{split}
 a(T) &= a_0 + a_1 \left(\frac{T_0}{T}\right)
 + a_2 \left(\frac{T_0}{T}\right)^2 , \\
 b(T) &= b_3\left(\frac{T_0}{T}\right)^3 \;.
 \end{split}
\label{eq:lp}
\end{equation}
The standard choice of the parameters reads~\cite{Ratti:2005jh};
\begin{equation}
 a_0 = 3.51\,, \quad a_1 = -2.47\,, \quad
 a_2 = 15.2\,, \quad b_3 = -1.75\,.
\end{equation}
The parameter $T_0$ in Eq.~\eqref{eq:Poly} sets the deconfinement
scale in the pure gauge theory, i.e.\ $T_c = 270\MeV$.

We assume a homogeneous magnetic field, $\bm B$, along the positive
$z$ axis.  The eigenvalues of the Dirac operator can be derived by the
Ritus method~\cite{Ritus:1972ky}, which are~\cite{Fukushima:2008xe};
\begin{equation}
 \omega_s^2 = M^2 + \bigl[ |\bm p| + s\,\mu_5 \text{sgn}(p_z) \bigr]^2 \;,
\label{eq:omega}
\end{equation}
apart from (the phases of) the Polyakov loop, where $s=\pm 1$,
$\bm p^2 = p_z^2 + 2|q_f B| k$ with $k$ a non-negative integer
labelling the Landau level.

The thermodynamic al potential $\Omega$ in the mean-field approximation
in the presence of an Abelian chromomagnetic field has been considered
in many literatures, Ref.~\cite{Campanelli:2009sc} for example.  The
expression for an electromagnetic $\bm B$ can be obtained in the same
way.  Here we simply write the final result;
\begin{align}
 & \Omega = \mathcal{U} + \frac{\sigma^2}{G}
  -N_c\sum_{f=u,d}\frac{|q_fB|}{2\pi} \sum_{s,k}\alpha_{sk}
  \int_{-\infty}^\infty \frac{d p_z}{2\pi} \,
  f_\Lambda^2 \, \omega_s(p) \notag \\
 &\quad -2T\sum_{f=u,d}\frac{|q_fB|}{2\pi} \sum_{s,k} \alpha_{sk}
  \int_{-\infty}^\infty \frac{d p_z}{2\pi} \notag\\
 &\qquad\times \ln\bigl( 1+3\,\Phi e^{-\beta \omega_s}
  + 3\,\bar\Phi e^{-2\beta \omega_s} + e^{-3\beta \omega_s} \bigr) \;.
\label{eq:O2}
\end{align}
Here the above definition of $\Omega$ is different from the standard
grand potential in thermodynamics by a volume factor $V$.  The
quasi-particle dispersion $\omega_s$ is given by Eq.~\eqref{eq:omega}.
The spin degeneracy factor is
\begin{equation}
 \alpha_{sk} = \left\{ \begin{array}{ll}
  \delta_{s,+1} & \text{   for~~~} k=0,~ qB>0 \;,\\
  \delta_{s,-1} & \text{   for~~~} k=0,~ qB<0 \;,\\
  1 & \text{   for~~~} k\neq0 \;.
 \end{array} \right.
\end{equation}
Before going ahead further, one may wonder why we introduce only one
order parameter for the chiral symmetry breaking even though the
magnetic field breaks isospin symmetry.  Since $q_u \neq q_d$, one
could suspect that the effects of $\bm B$ on $\langle\bar u u\rangle$
and on $\langle\bar d d\rangle$ are different.  This is not the case,
however, in the present model in the mean-field approximation.  As a
matter of fact, even in the presence of $\bm B\neq 0$, the
thermodynamic potential~\eqref{eq:O2} depends only on
$\sigma \propto \langle\bar u u\rangle + \langle\bar d d\rangle$.
This is so only when the four-fermion interaction is
Eq.~\eqref{eq:four-fermi} with equal mixing of the
$U(1)_A$-symmetric and $U(1)_A$-breaking terms.  Hence, the
relevant quantity for the chiral symmetry breaking is just one
condensate, namely $\sigma$, even for $\bm B \neq 0$, and there is no
need to introduce two independent condensates in this special case.
Even when we consider more general four-fermion interaction, the
isospin breaking effect is only negligibly small.

The vacuum part of the thermodynamic potential, $\Omega(T=0)$, is
ultraviolet divergent.  This divergence is transmitted to the gap
equations.  Thus we must specify a scheme to regularize this
divergence.  The choice of the regularization scheme is a part of the
model definition and, nevertheless, the physically meaningful results
should not depend on the regulator eventually.  In the case with a
strong magnetic field the sharp momentum cutoff suffers from cutoff
artifact since the continuum momentum is replaced by the discrete
Landau quantized one.  To avoid cutoff artifact, in this work, we use
a smooth regularization procedure by introducing a form factor
$f_\Lambda(p)$ in the diverging zero-point energy.  Our choice of
$f_\Lambda(p)$ is as follows;
\begin{equation}
 f_\Lambda(p) = \sqrt{\frac{\Lambda^{2N}}
  {\Lambda^{2N} + |\bm p|^{2N}}}\;,
\label{eq:f_p}
\end{equation}
where we specifically choose $N=10$.  In the $N\to\infty$ limit the
above $f_\Lambda(p)$ is reduced to the sharp cutoff function
$\theta(\Lambda-|\bm p|)$.  Since the thermal part of $\Omega$ is not
divergent, we do not need to introduce a regularization function.

\section{Phase structure with chiral chemical potential}
\label{sec:phase}

In this section we firstly focus on the system at $\mu_5=0$ and
discuss the role of the magnetic field as a catalyzer of the dynamical
chiral symmetry breaking.  We also analyze the interplay between
chiral symmetry restoration and deconfinement crossover as the
strength of $\bm B$ increases.

\subsection{Results at $\mu_5=0$
            -- chiral symmetry breaking and deconfinement}
We analyze, within the PNJL model, the response of quark matter to
$\bm B$ at $\mu_5 = 0$.  In particular we are interested in the
interplay between chiral symmetry restoration and deconfinement
crossover in the presence of a magnetic field which leads to the
so-called chiral magnetic catalysis~\cite{Gusynin:1995nb}.  Our model
parameter set is
\begin{equation}
 \Lambda = 620\MeV\;,\qquad
 G\Lambda^2 = 2.2\;.
\label{eq:param}
\end{equation}
These parameters correspond to $f_\pi = 92.4\MeV$ and the vacuum
chiral condensate $\langle\bar u u\rangle^{1/3} = -245.7\MeV$, and the
constituent quark mass $M=339\MeV$.  The critical temperature for
chiral restoration in the NJL part at $\bm B=0$ is
$T_c \approx 190\MeV$.  We set the deconfinement scale $T_0$ in the
Polyakov loop potential (see Eq.~\eqref{eq:Poly}) as $T_0 = 270\MeV$,
which is the value of the known deconfinement temperature in the pure
$SU(3)$ gauge theory.

\begin{figure}[t]
\begin{center}
\includegraphics[width=8.5cm]{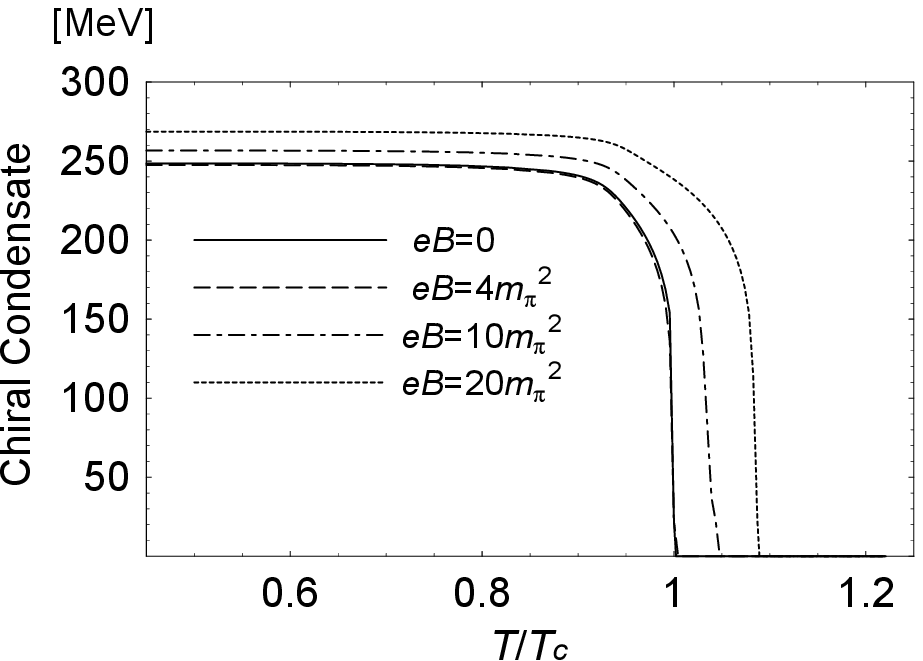}\vspace{1em}\\
\includegraphics[width=8.5cm]{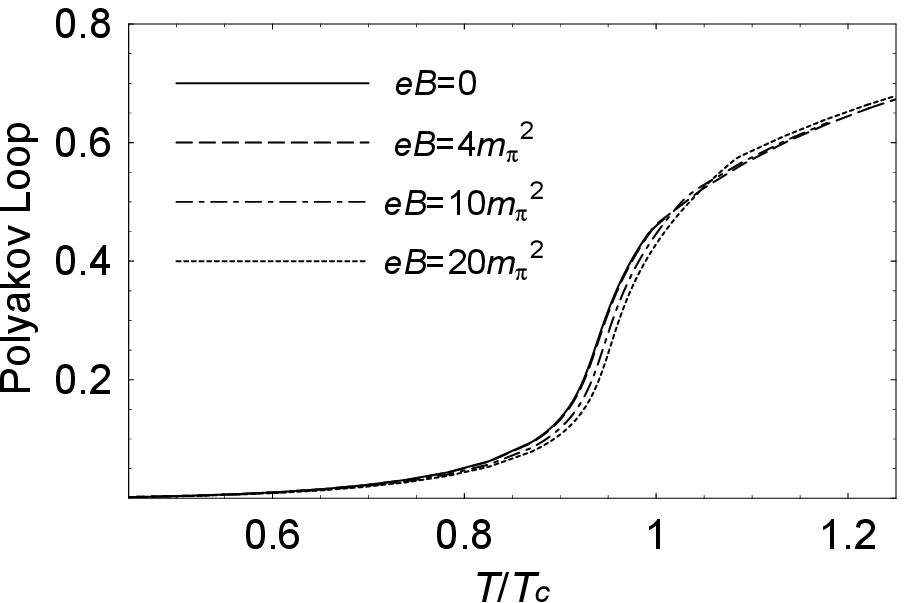}
\end{center}
 \caption{\label{FigS} Absolute value of the chiral condensate
 $|\langle\bar u u\rangle^{1/3}|$ (upper panel) and expectation value
   of the Polyakov loop (lower panel) as a function of $T$ computed at
   several values of $eB$ (in unit of $m_\pi^2$).  In this model
   $T_c=228\MeV$ at $\mu_5=B=0$.}
\end{figure}

In Fig.~\ref{FigS} we plot the absolute value of the chiral condensate
$\langle\bar u u\rangle^{1/3}$ (upper panel) and expectation value of
the Polyakov loop (lower panel) as a function of $T$ computed at
several values of $eB$ (expressed in unit of $m_\pi^2$).  The chiral
condensate $\langle\bar u u\rangle$ and the Polyakov loop $\Phi$ are
the solution of the gap equations
$\partial\Omega/\partial\sigma=\partial\Omega/\partial\Phi=0$ in the
model at hand.

Figure~\ref{FigS} is interesting for several reasons.  First of all,
the role of $\bm B$ as a catalyzer of chiral symmetry breaking is
evident.  Indeed, the chiral condensate and thus constituent quark
mass increase in the whole $T$ region as $eB$ is raised (for
graphical reasons, we have plotted our results starting from
$T=100\MeV$.  There is nevertheless no significant numerical
difference between the $T=0$ and $T=100\MeV$ results).  This behavior
is in the correct direction consistent with the well-known magnetic
catalysis revealed in Ref.~\cite{Gusynin:1995nb} and also discussed
recently in Ref.~\cite{Campanelli:2009sc} in a slightly different
context of the PNJL-model study on the response of quark matter to
external chromomagnetic fields.

Secondly, we observe that the deconfinement crossover is only
marginally affected by the magnetic field.  We can identify the
deconfinement $T_c$ by the inflection point of $\Phi$ as a function of
$T$.  This simple procedure gives results nearly in agreement with
those obtained by the peak position in the Polyakov loop
susceptibility, which is a common prescription to locate the so-called
pseudo-critical temperature.  Also we can identify the deconfinement
$T_c$ with the temperature at which $\Phi=0.5$.  We note that $T_c$ in
Fig.~\ref{FigS} is the chiral $T_c=228\MeV$ where the chiral
condensate vanishes, but not the deconfinement $T_c$ in both figures.

From Fig.~\ref{FigS} we notice that, increasing $eB$ from $4m_\pi^2$
to $20m_\pi^2$, the shift of the chiral transition temperature
$\Delta T_\chi \approx 20\MeV$, while the shift of the deconfinement
crossover temperature is as small as $\Delta T_\Phi \approx 5\MeV$.
Hence, the chiral phase transition is more easily influenced by the
magnetic field than the deconfinement as anticipated.  Consequently,
under a strong magnetic field, there opens a substantially wide
$T$-window in which quarks are deconfined but chiral symmetry is still
spontaneously broken.  This result, valid for $\mu_5= 0$, does not
necessarily hold, in general, for $\mu_5\neq 0$, as we will see in the
next subsection.

\begin{figure*}
\begin{center}
\includegraphics[width=8.5cm]{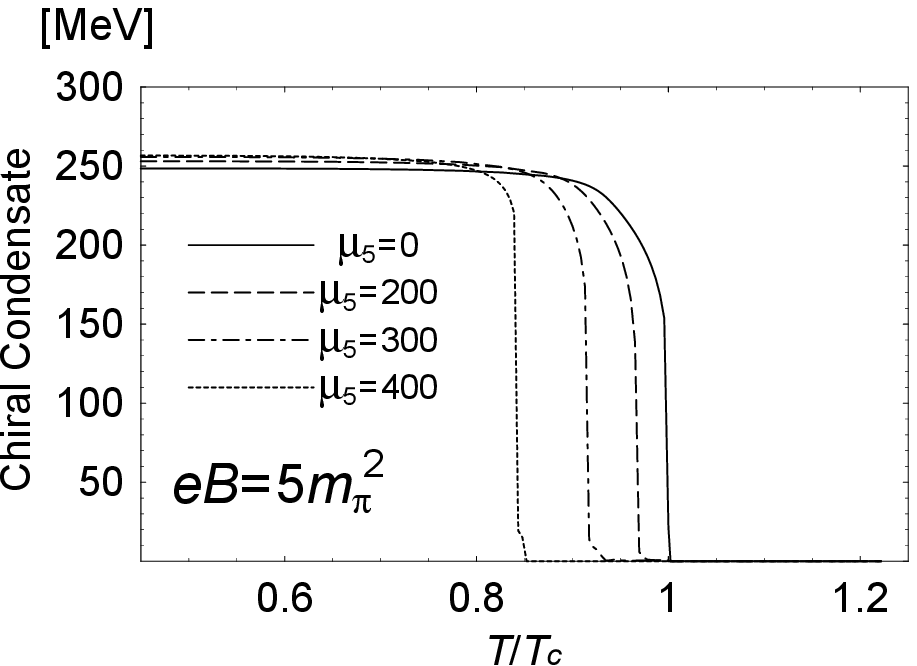}~~
\includegraphics[width=8.5cm]{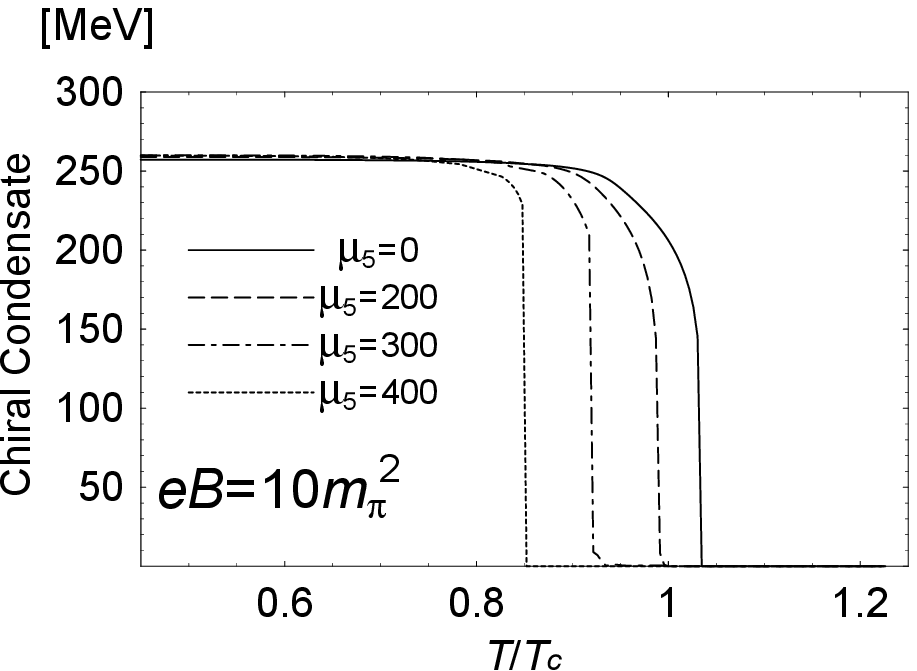} \vspace{1em}\\
\includegraphics[width=8.5cm]{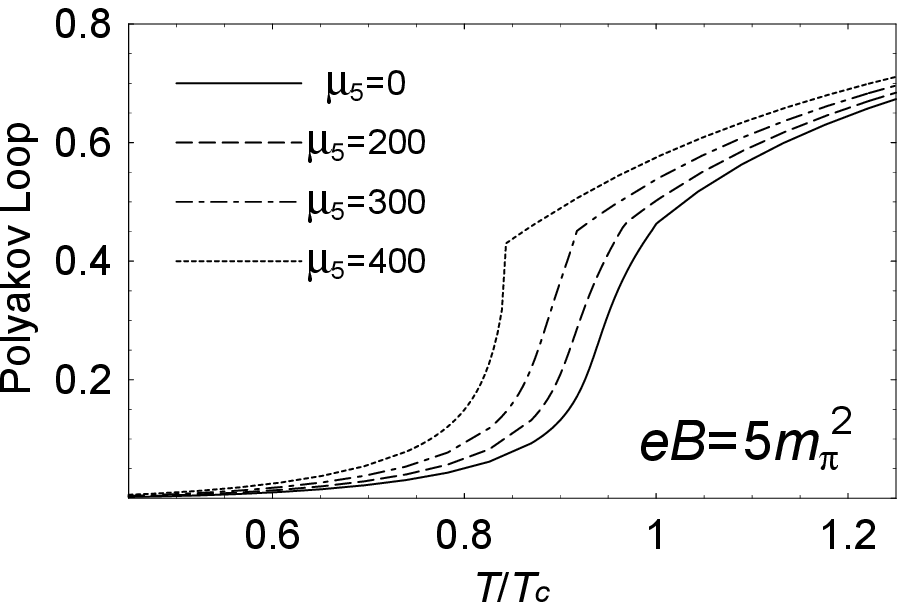}~~
\includegraphics[width=8.5cm]{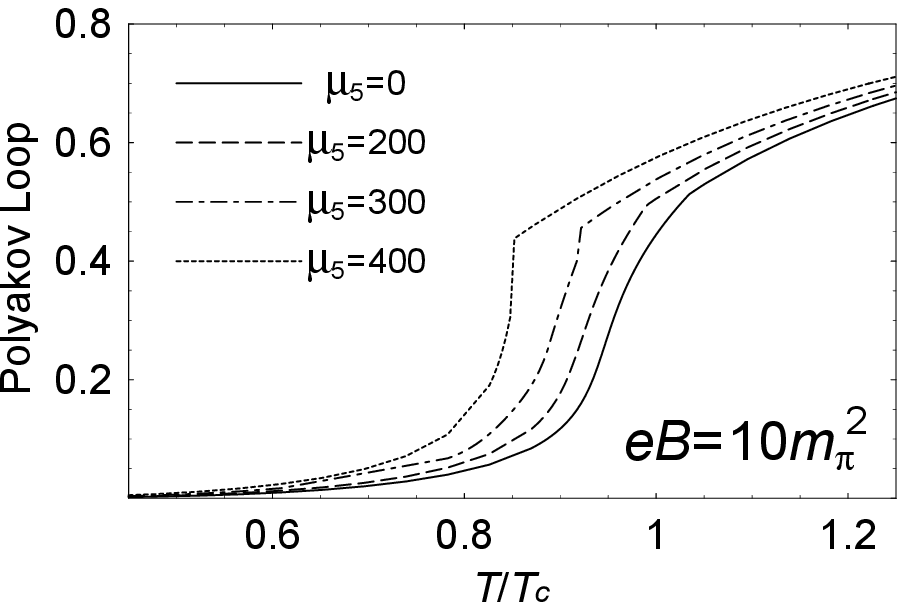}
\end{center}
\caption{\label{Fig:Masses} Absolute value of the chiral condensate
  $|\langle\bar u u\rangle^{1/3}|$ (upper panel) and expectation value
  of the Polyakov loop (lower panel) as a function of $T$ computed at
  several values of $eB$ (in unit of $m_\pi^2$) and $\mu_5$ (in unit
  of MeV).}
\end{figure*}

\subsection{Results at $\mu_5\neq0$
            -- suppression on the chiral condensate}
We next turn to the study of the effect of a finite $\mu_5$ on the QCD
phase transitions using the PNJL model.
We recall that $\mu_5$ cannot be a true chemical
  potential since its conjugate variable $n_5$ is only approximately
  conserved due to axial anomaly.  Nevertheless, because the time
  derivative of the strong $\theta$ angle translates into $\mu_5$, as
  explained in the introduction section, $\mu_5$ itself is a
  physically meaningful quantity.
We specifically look into the behavior of the critical line for chiral
symmetry restoration, which is well defined in the chiral limit, at
differing $\bm B$ while keeping $\mu_5$ fixed.  This study will be
useful, among other things, in order to understand the relation
between the chirality density $n_5$ and $\mu_5$ that we compute
numerically in later discussions.

One effect of $\mu_5\neq0$ is lowering of the critical temperature of
the chiral phase transition.  This is evident from the upper panels of
Fig.~\ref{Fig:Masses}.  Firstly we discuss the case of $eB = 5m_\pi^2$
corresponding to the left upper and left lower panels of
Fig.~\ref{Fig:Masses}.  We see that increasing $\mu_5$ at low $T$
results in slight enhancement of the chiral condensate.  As $T$
approaches $T_c$, however, the chiral phase transition at larger
$\mu_5$ takes earlier place below $T_c$.  As a result of the coupling
between the chiral condensate and the Polyakov loop, the deconfinement
crossover as shown in the lower panels of Fig.~\ref{Fig:Masses} is
also shifted earlier as $\mu_5$ becomes greater.

In view of the right upper and right lower panels of
Fig.~\ref{Fig:Masses} for large $eB = 10m_\pi^2$ the $\mu_5$-effect on
the chiral condensate at low $T$ is less visible.  This is understood
from the fact that the chiral magnetic catalysis effect is predominant
over the minor enhancement due to $\mu_5$.  In contrast, as $T$ is
increased toward $T_c$, the qualitative behavior of the shift in the
critical temperature is just the same as what we have seen previously
for $eB=5m_\pi^2$.

An interesting prediction from the PNJL model is that, at a given
value of $eB$, there exists a critical $\mu_5$, above which the chiral
phase transition becomes first order.  In the case $eB=5m_\pi^2$ as
shown in Fig.~\ref{Fig:Masses} the critical $\mu_5$ is found between
$300\sim 400\MeV$.  As a matter of fact, at $\mu_5 = 400\MeV$, the
chiral condensate and the Polyakov loop both exhibit discontinuity at
the critical temperature.  We see that, as compared to the
$\mu_5 = 300\MeV$ case, the slopes of the chiral condensate and the
Polyakov loop sharply change as a function of $T$ for the
$\mu_5=400\MeV$ case.  Hence, our data plotted in
Fig.~\ref{Fig:Masses} suggest the existence of a critical $\mu_5$ in
the range $300\MeV<\mu_5^c<400\MeV$ at which the weakly first-order
transition becomes a true second-order one.  The phase diagram in the
$\mu_5$-$T$ plane has a tricritical point (TCP) accordingly.  We will
discuss more on the TCP in the next subsection.  We notice that this
picture is qualitatively robust regardless of the chosen value of
$eB$, as is already implied from Fig.~\ref{Fig:Masses}.  The
first-order phase transition in the high-$mu_5$ and low $T$ region
leads to a discontinuity in the chirality density as a function of
$\mu_5$.  This point will be also addressed in some details in the
next section.

As a final remark in this subsection we note that, in
  Figs.~\ref{FigS} and \ref{Fig:Masses}, the slope of the curve
  quickly changes at some point (at $T/T_c=1$ for $eB=0$ for example).
  This is because of the second-order phase transition which is the
  case for chiral restoration in the chiral limit.

\subsection{Phase diagram}
\begin{figure}
\begin{center}
\includegraphics[width=8.5cm]{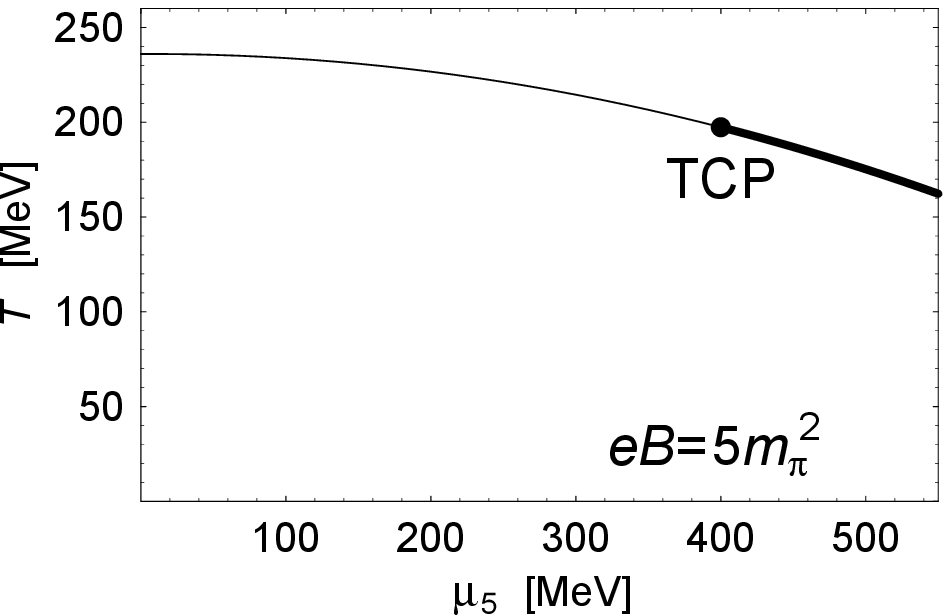} \vspace{1em}\\
\includegraphics[width=8.5cm]{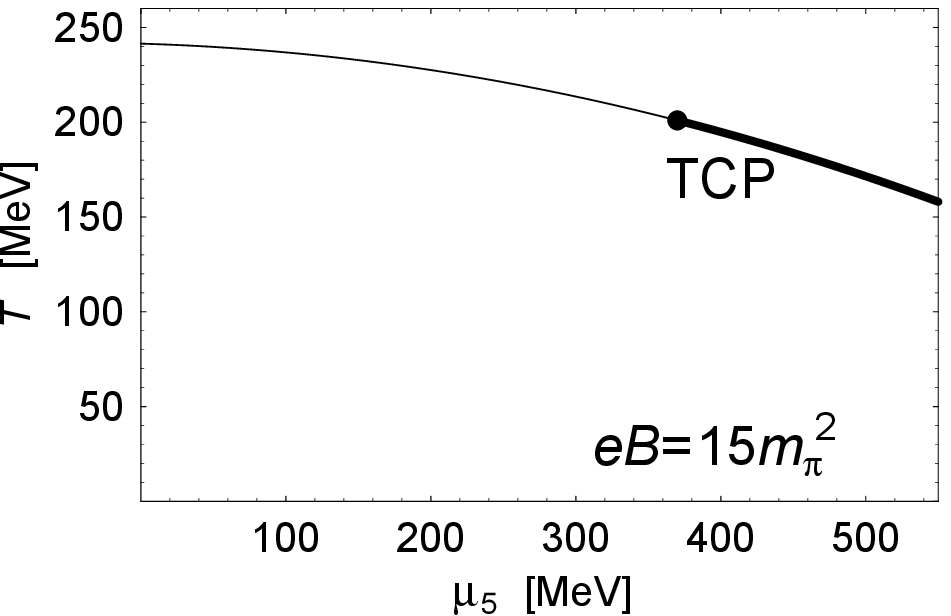}
\end{center}
\caption{\label{Fig:thePD} (Upper panel) Phase diagram in the
$\mu_5$-$T$ plane obtained at $eB=5 m_\pi^2$.  The thin line
  represents a second-order chiral phase transition and the thick
  one a first-order transition.  Below the line, chiral symmetry is
  spontaneously broken, while chiral symmetry is restored above the
  line.  The label ``TCP'' denotes the tricritical point.  (Lower
  panel) Phase diagram for $eB=15 m_\pi^2$.}
\end{figure}

The results we have revealed so far can be summarized into the phase
diagram in the $\mu_5$-$T$ plane.  In the upper panel of
Fig.~\ref{Fig:thePD} we show the phase diagram at $eB=5m_\pi^2$.  In
the lower panel for comparison we plot the phase diagram at
$eB=15m_\pi^2$.  The line represents the chiral phase transition.  It
is of second order for small values of $\mu_5$ (shown by a thin line)
and becomes of first order at large $\mu_5$ (shown by a thick line).
The location of the TCP on the phase diagram depends only slightly on
$eB$, while the topology of the phase diagram is not sensitive to the
magnetic field.

The general effect of $\mu_5$ is to lower the chiral transition
temperature.  One may argue that the critical line can hit $T=0$
eventually at very large $\mu_5$, though the PNJL model is of no use
at such large $\mu_5$ because the ultraviolet cutoff causes unphysical
artifacts.  The locations of the TCP are estimated from the PNJL model
as
\begin{align}
 \left(\mu_5, T\right) &\approx (400\MeV,\; 200\MeV) \;,
  \text{ for } eB=5m_\pi^2 \;,\\
 \left(\mu_5, T\right) &\approx (370\MeV,\; 200\MeV) \;,
  \text{ for } eB=15m_\pi^2 \;.
\end{align}

\section{Chirality charge, electric current, and susceptibilities}
\label{sec:cme}

In this section we show our results for quantities relevant to the
Chiral Magnetic Effect (CME).  We numerically compute the chiral
density $n_5$ and its susceptibility $\chi_5$ as a function of $\mu_5$
and $eB$.  Also we calculate the current density $j_3$ along the
direction of $\bm B$ and its susceptibility $\chi_J$.  Finally, we use
the result $n_5(\mu_5)$ to evaluate $j_3$ as a function of $n_5$.

\subsection{Chirality density and its susceptibility}

\begin{figure}
\begin{center}
\includegraphics[width=8cm]{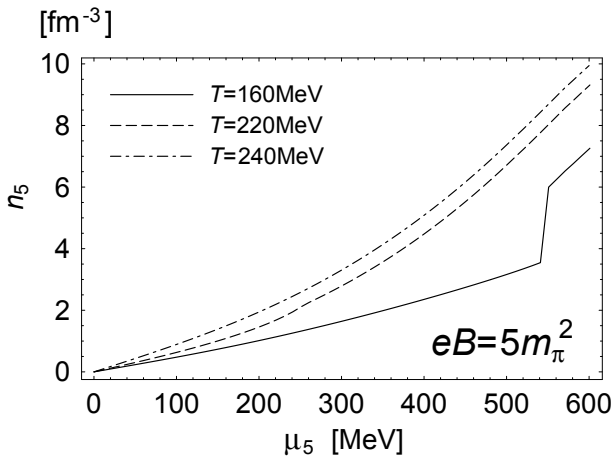}
\end{center}
\caption{\label{Fig:Current02} Chirality density $n_5$ (in unit of
  fm$^{-3}$) as a function of $\mu_5$ (in unit of MeV) at
  $eB=5m_\pi^2$ for several values of $T$.}
\end{figure}

The axial anomaly relates the topological charge $Q_W$ to the
chirality charge $N_5$ with $N_5 = n_5 V$ where $V=L_x L_y L_z$ is the
volume of topological domains.  We can read $n_5$ from
\begin{equation}
 n_5 = -\frac{\partial\Omega}{\partial\mu_5} \;.
\end{equation}
It is useful information to relate $n_5$ and $\mu_5$ for various
temperatures and magnetic field strength.  In the next section we will
use the results of $n_5(\mu_5)$ to express the current density as a
function of the chirality density.

The relation between $n_5$ and $\mu_5$ can be found analytically only
in simple limiting cases~\cite{Fukushima:2008xe}.  In general one has
to determine it numerically using an effective model.  We show
$n_5(\mu_5)$ for $eB=5m_\pi^2$ at three temperatures around $T_c$ in
Fig.~\ref{Fig:Current02}.    The qualitative picture is hardly modified even
if we change the magnetic field.

From Fig.~\ref{Fig:thePD} we can read the critical temperature at
$\mu_5 = 0$ that is $T_c=228\MeV$.  At temperatures well below $T_c$,
as seen in the $T=160\MeV$ case in Fig.~\ref{Fig:Current02}, the
discontinuity associated with the first-order phase transition with
respect to $\langle\bar{u}u\rangle^{1/3}$ and $\Phi$ is conveyed to
the relation $n_5(\mu_5)$, which is a typical manifestation of the
mixed phase at critical $\mu_5$.  Naturally, as $T$ gets larger, the
chirality density as a function of $\mu_5$ becomes smoother, since the
chiral phase transition is of second order at higher $T$ as is clear
from Fig.~\ref{Fig:thePD}.

\begin{figure*}[t]
\begin{center}
 \includegraphics[width=8cm]{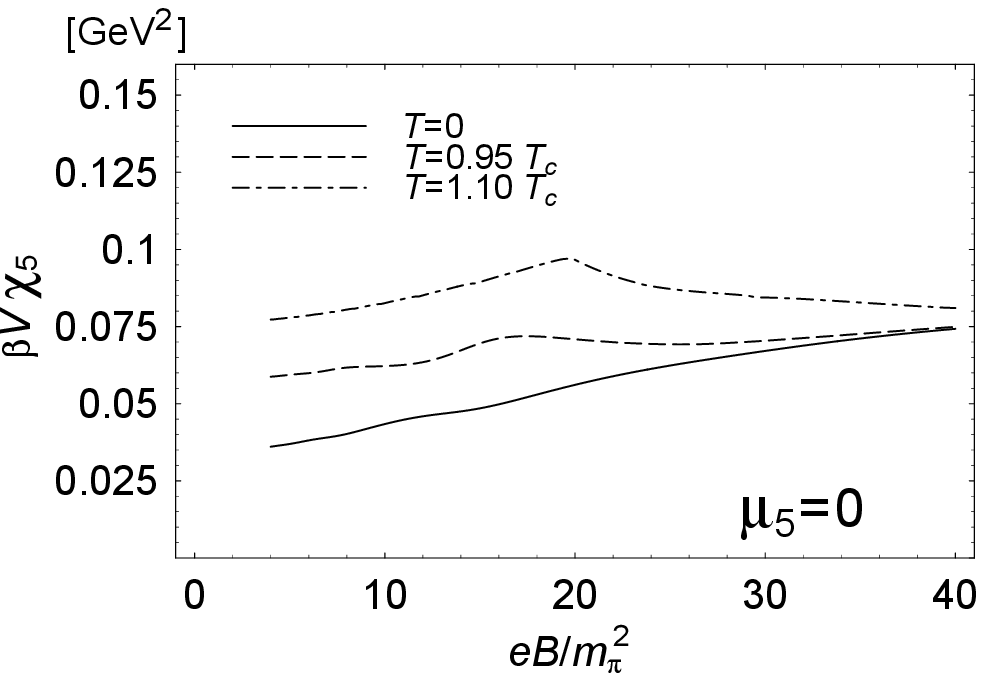}~~
 \includegraphics[width=8cm]{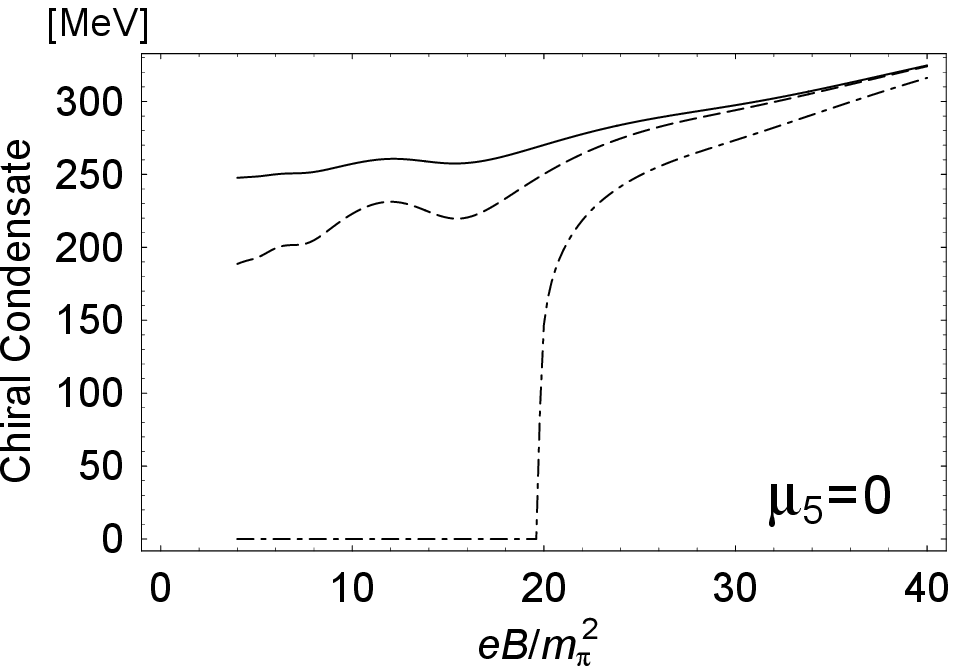} \vspace{1em}\\
 \includegraphics[width=8cm]{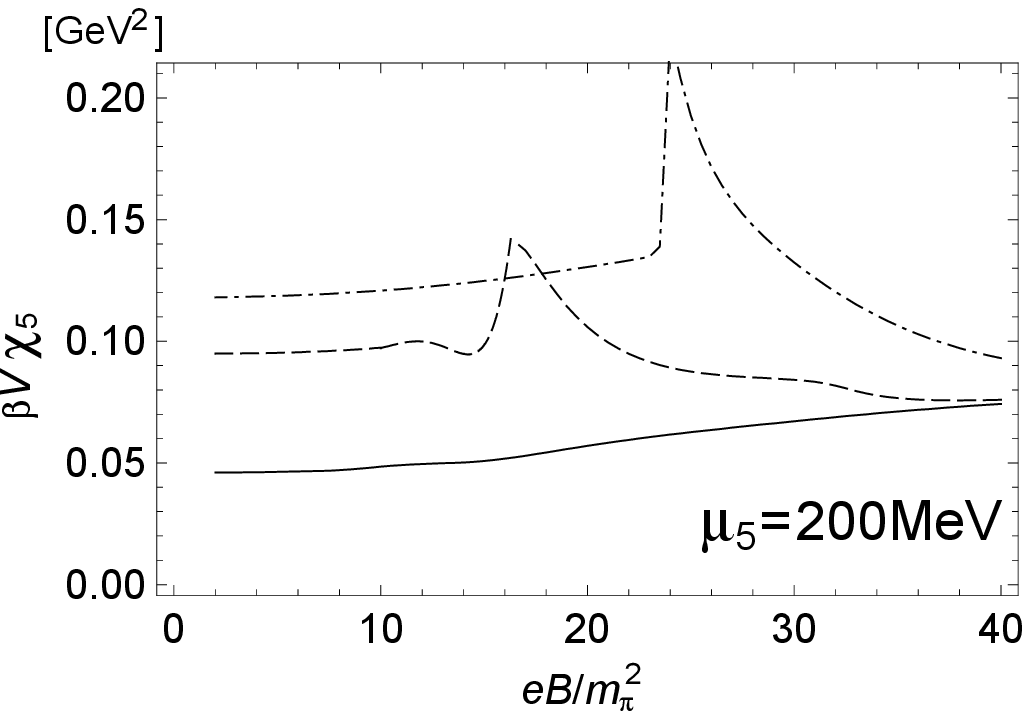}~~
 \includegraphics[width=8cm]{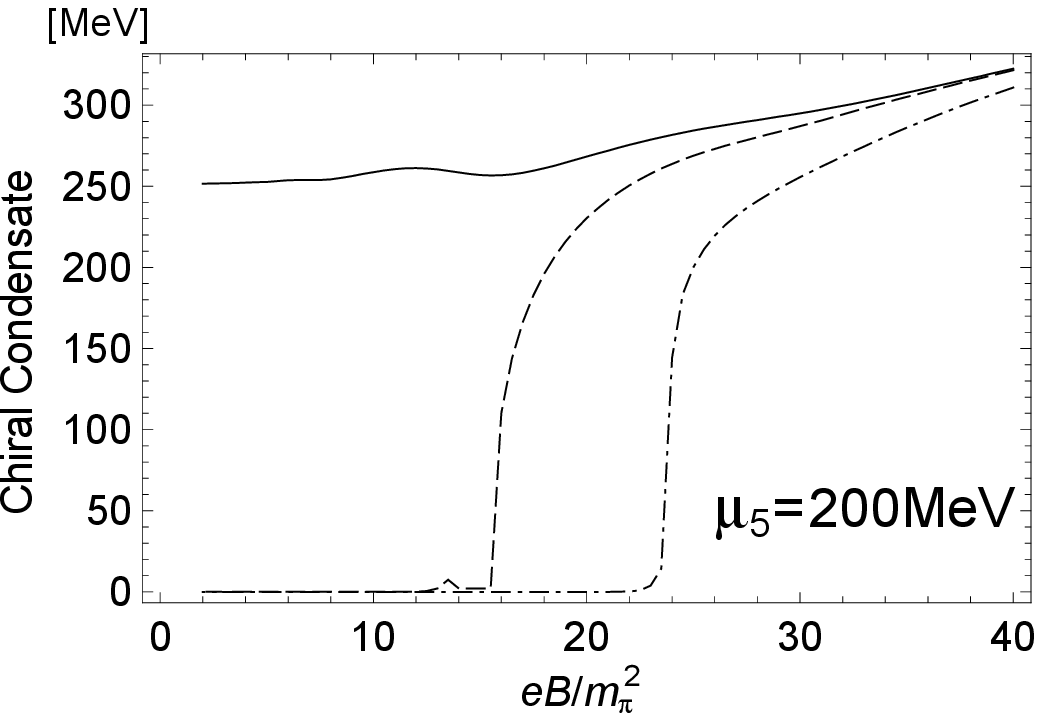} \vspace{1em}\\
 \includegraphics[width=8cm]{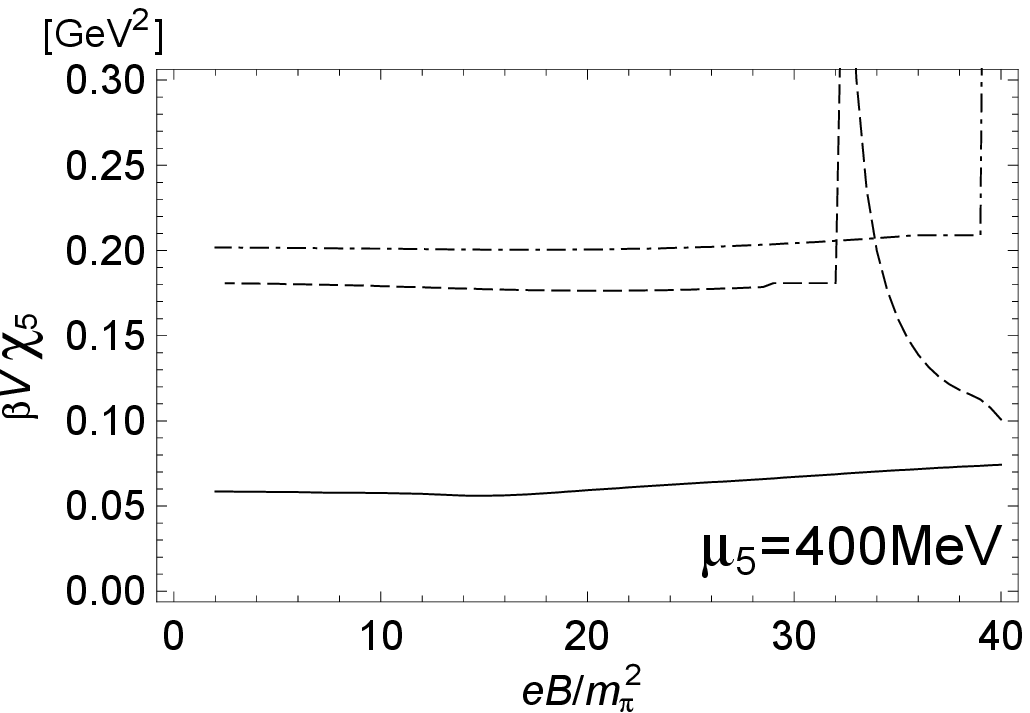}~~
 \includegraphics[width=8cm]{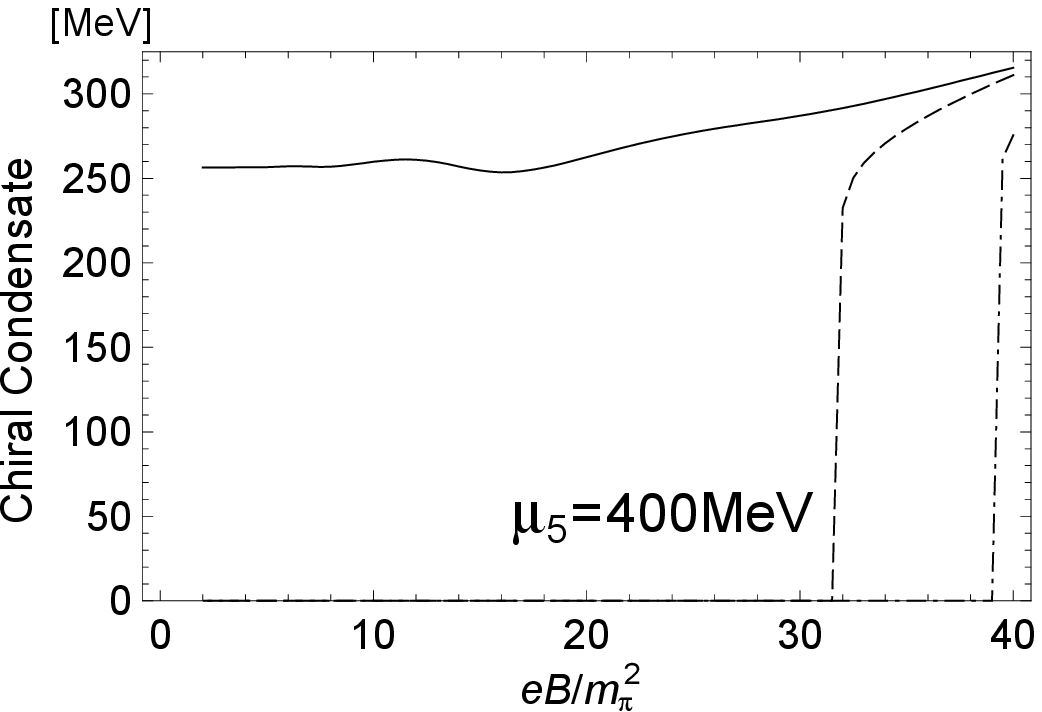}
\end{center}
\caption{\label{Fig:Susc02} (Left panels) Chirality charge
  susceptibility $\chi_5$ as a function of $eB$ (in unit of $m_\pi^2$)
  for several temperatures.  The chiral chemical potential is chosen
  as $\mu_5=0$, $200\MeV$, and $400\MeV$, respectively, from the upper
  to the lower panels.  The solid line corresponds to $T=0$, the
  dashed line $T=0.95T_c$, and the dot-dashed line $T=1.1 T_c$.  Here
  $T_c=228\MeV$ is the critical temperature in this model at
  $\mu_5=B=0$.  (Right panels) Absolute value of the chiral condensate
  $\langle\bar u u\rangle^{1/3}$ as a function of $eB$.  The line
  styles are the same defined in the left upper panel.}
\end{figure*}

It is interesting to compute the chirality charge susceptibility
$\chi_5$, as well as $n_5$, defined as
\begin{equation}
 \chi_5 = \langle n_5^2\rangle - \langle n_5\rangle^2
  = -\frac{1}{\beta V}\frac{\partial^2\Omega}{\partial\mu_5^2}\;,
\label{eq:chi5}
\end{equation}
where $\beta = 1/T$ and $V$ the volume.  We note that this definition
of the susceptibility is different from that in
Ref.~\cite{Fukushima:2009ft} by $V$.  It should be mentioned that we take a
numerical derivative to compute $\chi_5$ including implicit dependence in
$\Phi$ and $\sigma$. In Fig.~\ref{Fig:Susc02} we plot
$\chi_5$ as a function of $eB$ for several $\mu_5$ values.  The upper
panel corresponds to $\mu_5=0$, middle one to $\mu_5 = 200\MeV$, and
lower one to $\mu_5 = 400\MeV$.  For completeness, in the right panels
of the same figure~\ref{Fig:Susc02} we plot the chiral condensate
$|\langle\bar u u\rangle^{1/3}|$ for the same $T$ and same $\mu_5$.
The oscillations in $\chi_5$ are artificial results because of the
momentum cutoff $\Lambda$.  As shown in Ref.~\cite{Campanelli:2009sc},
choosing a regulator which is smoother than used in this work, the
oscillations of the various quantities could be erased.  The
qualitative picture is, nevertheless, unchanged even with a different
regulator.  For this reason we do not perform a systematic study here
on the cutoff scheme dependence.

A notable aspect is the suppression of the chirality-charge
fluctuations at large $T$ and large $eB$.  This is evident, for
example, in the result with $\mu_5=0$ and $T=1.1T_c$ in
Fig.~\ref{Fig:Susc02}.  As long as $eB$ is small, $\chi_5$ is a
monotonously increasing function of $eB$ as expected naively.  When
$eB$ reaches a critical value around $20m_\pi^2$, however, $\chi_5$
has a pronounced peak and then decreases with increasing $eB$, which
is a result of mixture with diverging chiral susceptibility at the
chiral phase transition.  It should be mentioned that $\chi_5$ at
$\mu_5=0$ (as shown in the upper left panel of Fig.~\ref{Fig:Susc02})
does not diverge at the critical $eB$ since the mixing with the chiral
susceptibility is vanishing due to $\mu_5=0$.  This behavior below and
above the chiral critical point can be easily understood in terms of
the chiral symmetry breaking by virtue of the magnetic field.  As a
matter of fact, at $T>T_c$ the chiral condensate stays zero
identically as long as $eB$ is small enough, leading to zero
quasiparticle masses.  Once $eB$ exceeds a critical value, the chiral
symmetry is broken spontaneously even at high $T$ (see the upper right
panel of Fig.~\ref{Fig:Susc02}) and the quasiparticle masses can then
jump to a substantially large number then.  Such dynamical quark
masses result in appreciable suppression of the chirality-charge
fluctuations.  As it will be shown in the next section, this
interesting and intuitively understandable effect appears in the
current susceptibility as well.

\subsection{Current density and its susceptibility}
The current density $j_3$ (and its susceptibility as well) is the most
important quantity to compute for the Chiral Magnetic
Effect~\cite{Fukushima:2009ft}.  It corresponds to the charge per unit
volume that moves in the direction of the applied magnetic field in a
domain where an instanton/sphaleron transition takes place, which
causes chirality change of quarks.  The current has been computed
analytically in Ref.~\cite{Fukushima:2008xe} in four different ways.

To compute the current density along the magnetic field, i.e.\
$j_3 = q\langle\bar\psi\gamma_3\psi\rangle$, we follow the common
procedure to add an external homogeneous vector potential along the
magnetic field, $A_3$, coupled to the fermion field.  Then,
\begin{equation}
 j_3 = -\frac{\partial\Omega}{\partial A_3}\biggr|_{A_3=0}\;.
\label{eq:numDer}
\end{equation}
The derivative of the thermodynamic potential in the presence of a
background field is computed in the following way.  The coupling of
quarks to $A_3$ is achieved by shifting $p_z$ in Eq.~\eqref{eq:O2} as
$p_z\rightarrow p_z + q_f A_3$.  After putting regularization in the
momentum integral with an ultraviolet cutoff $\Lambda$ (we know that
the current is ultraviolet finite, hence the choice of the
regularization method does not affect the final result) we change the
order of the momentum integral and the derivative with respect to
$A_3$.  Then we make use of the following replacement,
\begin{equation}
 \frac{\partial}{\partial A_3} \to q_f \frac{d}{d p_z}\;,
\label{eq:der2}
\end{equation}
to obtain,
\begin{equation}
 j_3 = N_c \sum_{f=u,d} q_f\frac{|q_f B|}{2\pi} \sum_{s,k}\alpha_{ks}
  \int_{-\Lambda}^{\Lambda}\frac{dp_z}{2\pi}\frac{d}{dp_z}
  \left[  \omega_s(p) + \cdots \right] \;.
\end{equation}
The ellipsis represents irrelevant matter terms.  After summing over
the spin $s$, the contribution of the integrand from the Landau levels
with $n>0$ turns out to be an odd function of $p_z$.  Therefore, only
the lowest Landau level gives a non-vanishing contribution to the
current and we get from the surface
contribution~\cite{Fukushima:2008xe},
\begin{equation}
 j_3 = N_c\sum_{f=u,d}\frac{q_f^2\mu_5 B}{2\pi^2}
  = \frac{5\mu_5 e^2B}{6\pi^2} \;,
\label{eq:J3one}
\end{equation}
which is certainly ultraviolet finite as it should be.
Generally speaking we should utilize a gauge-invariant
  regularization.  Nevertheless the above \eqref{eq:J3one} indicates
  that a naive momentum cutoff can reproduce a correct expression for
  the anomalous chiral magnetic current.

The current density as given by Eq.~\eqref{eq:J3one} does not depend
on quark mass explicitly, and on temperature either.  The reason is
that the current is generated by the axial anomaly and it receives
contributions only from the ultraviolet momentum regions (as the above
derivation shows), and so it is insensitive to any infrared energy
scales.  Also, the Polyakov loop does not appear explicitly in
Eq.~\eqref{eq:J3one}.  This is easy to understand; the Polyakov loop
is a thermal coupling between quark excitations and the gluonic
medium, and thus the Polyakov loop only enters the thermal part of
$\Omega$.  Since the current originates from the anomaly, however, the
thermal part of $\Omega$ just drops off for the current generation.
The effect of the Polyakov loop will appear implicitly through the
relation between $\mu_5$ to $n_5$.

\begin{figure}
\begin{center}
\includegraphics[width=8.5cm]{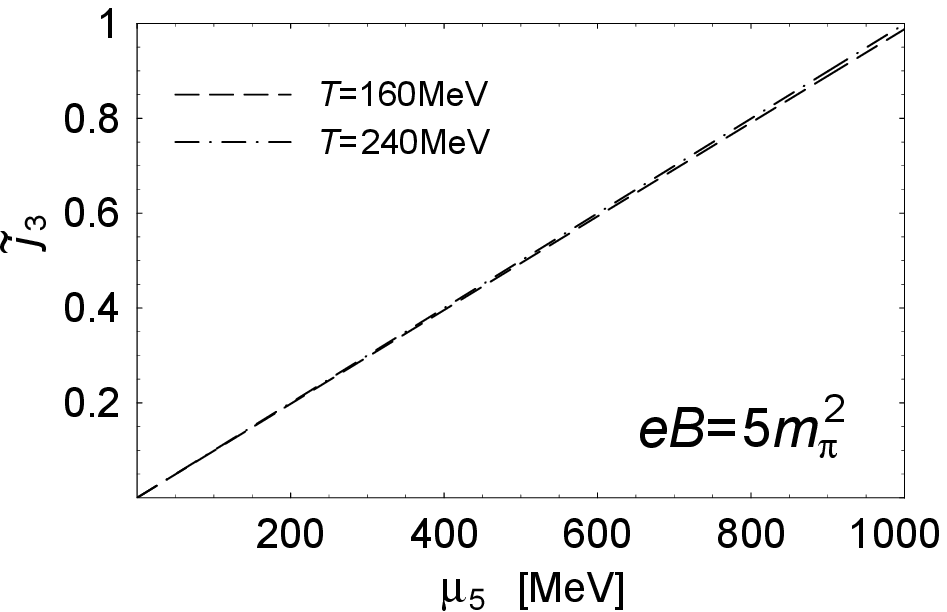} \vspace{2em}\\
\includegraphics[width=8.5cm]{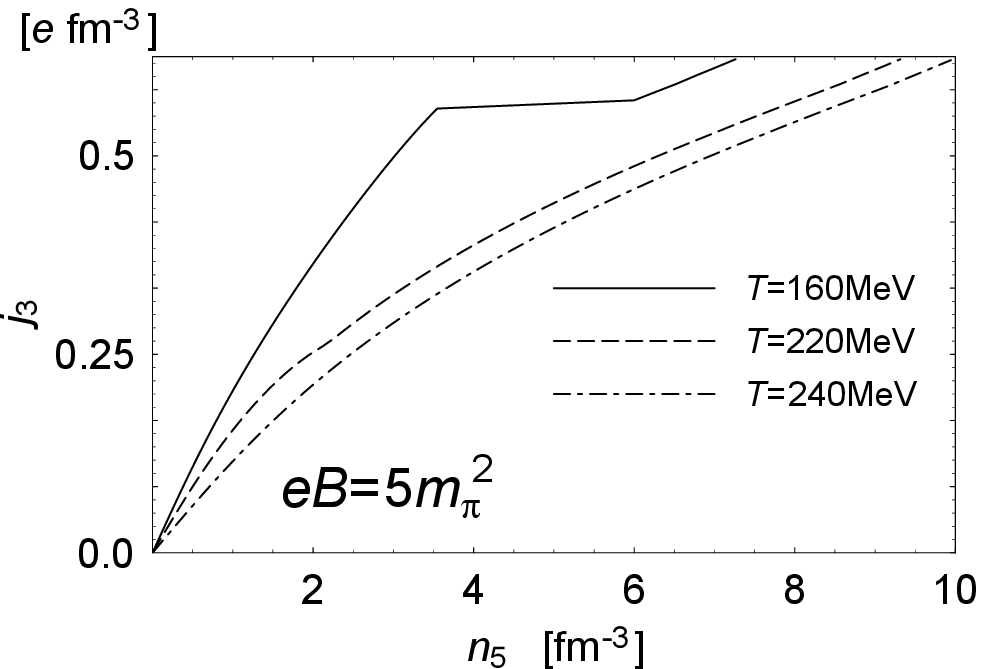}
\end{center}
\caption{\label{Fig:comparison1} (Upper panel) Normalized current
density, $\tilde{j}_3 = (5\mu_0\, e^2B/6\pi^2)^{-1} j_3$ with
$\mu_0=1\GeV$, as a function of $\mu_5$ at two different temperatures
(below and above $T_c$).  (Lower panel) Current density as a function
of $n_5$ for $eB=5m_\pi^2$ computed at three different temperatures.}
\end{figure}

To confirm that our numerical prescription works well, we have
computed $j_3$ by means of Eq.~\eqref{eq:numDer} with $\Omega$ given
in Eq.~\eqref{eq:O2}.  In Fig.~\ref{Fig:comparison1} we show the
results from our numerical computation as a function of $\mu_5$.
In the figure we have plotted the normalized current,
\begin{equation}
 \tilde{j}_3 = \biggl(\frac{5\mu_0\,e^2B}{6\pi^2}\biggr)^{-1} j_3
 \;,\label{eq:J3J3}
\end{equation}
with a choice of $\mu_0=1\;\text{GeV}$, which we defined so to make
the comparison transparent at a glance.  In Fig.~\ref{Fig:comparison1}
the dashed line represents $\tilde{j}_3$ at $T=160\MeV$; on the other
hand, the dot-dashed line the case at $T=240\MeV$.  We notice that our
numerical results are perfectly in agreement with
Eq.~\eqref{eq:J3one}.  We conclude that our numerical procedure
correctly reproduces the expected dependence of $j_3$ on $\mu_5$ with
the correct coefficient insensitive to infrared scales regardless of
whether $T$ is below or above $T_c$.

The result shown in the upper panel of Fig.~\ref{Fig:comparison1}
gives us confidence in our numerical procedure but the figure itself
is not yet more informative than Eq.~\eqref{eq:J3one}.  We express now
$j_3$ as a function of $n_5$ using Eq.~\eqref{eq:J3one} and the
results discussed in the previous section.  The result of this
computation is shown in the lower panel of Fig.~\ref{Fig:comparison1},
in which we plot the (not normalized) current density (measured in
fm$^{-3}$) as a function of $n_5$ (measured in fm$^{-3}$), at
$eB=5m_\pi^2$ and at three different temperatures.

From Fig.~\ref{Fig:comparison1} we notice that, at a fixed value of
$n_5$, the larger the temperature is, the smaller $j_3$ becomes.  This
seemingly counter intuitive result is easy to understand.  As a matter
of fact, as the temperature gets larger, the corresponding $\mu_5$ for
a given $n_5$ should decrease because of more abundant thermal
particles at higher temperature.  Since $j_3$ depends solely on
$\mu_5$, a higher temperature requires a larger $n_5$ to give the same
$j_3$.

\begin{figure}
\begin{center}
\includegraphics[width=8cm]{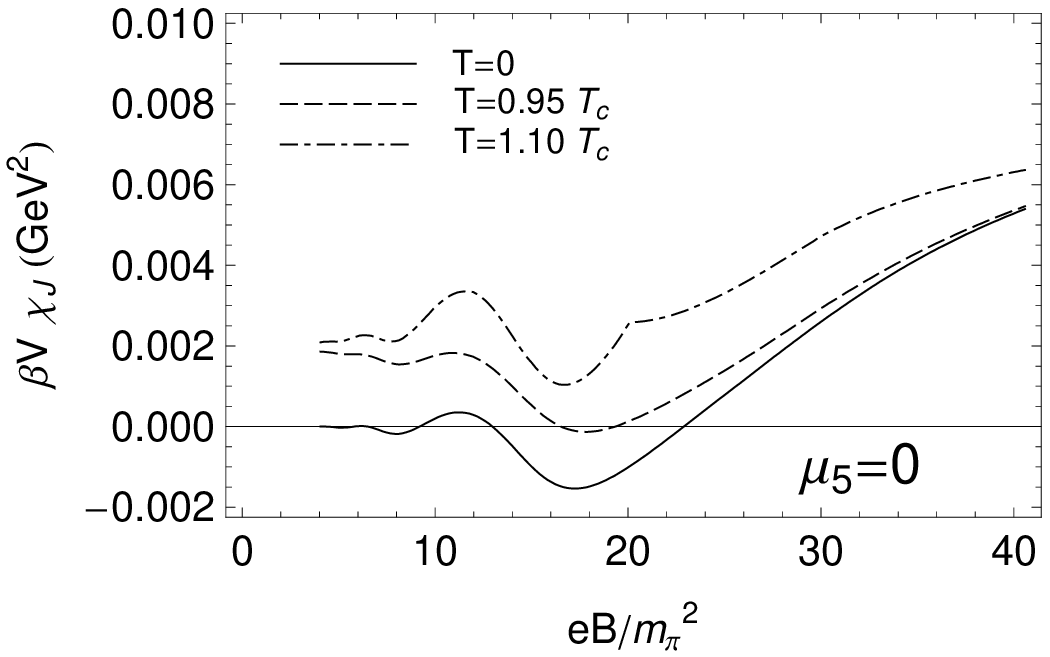} \vspace{1em}\\
\includegraphics[width=8cm]{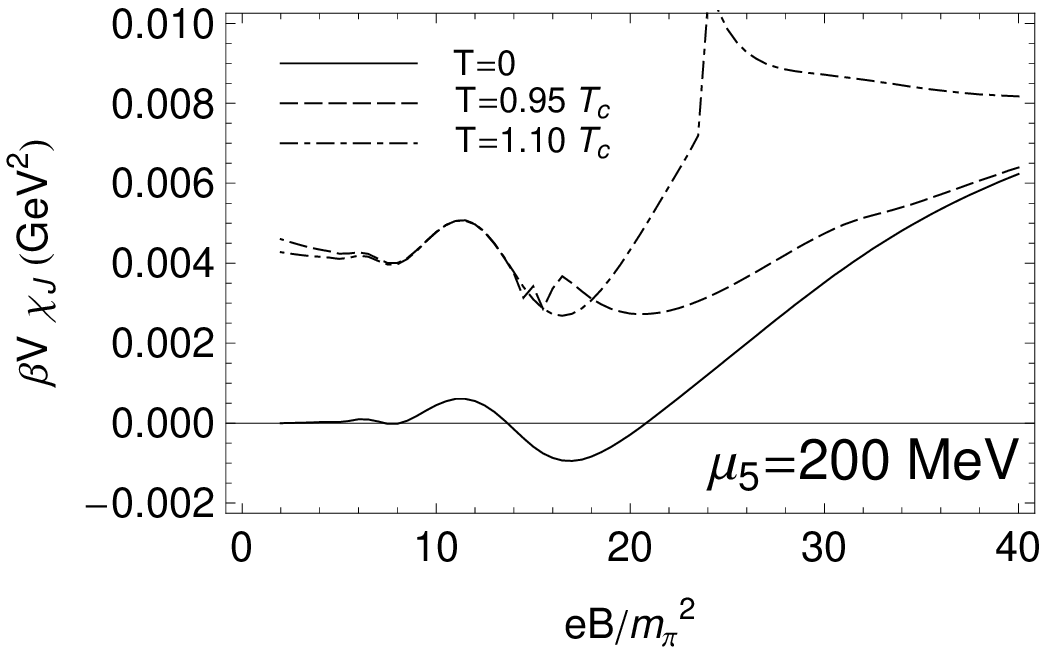} \vspace{1em}\\
\includegraphics[width=8cm]{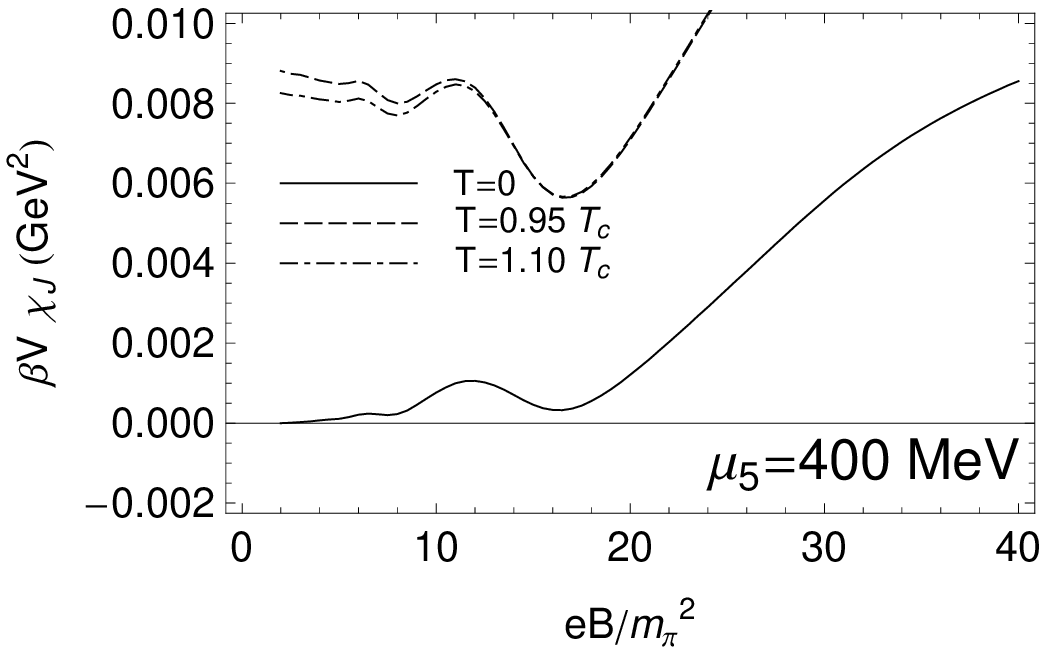}
\end{center}
\caption{\label{Fig:CurrentChi} Subtracted current susceptibility,
$\beta V\bar{\chi}_J$, as a function of $eB$ (in unit of $m_\pi^2$) for
several different values of $T$ (in unit of $T_c=228\MeV$) and $\mu_5$
(measured in MeV).}
\end{figure}

Besides $j_3$, another interesting quantity is the current
susceptibility defined by
\begin{equation}
 \chi_J = \langle j_3^2\rangle - \langle j_3\rangle^2
  = -\frac{1}{\beta V}\frac{\partial^2\Omega}{\partial A_3^2}
  \biggr|_{A_3 = 0} \;.
\label{eq:chi3}
\end{equation}
If we naively use the above definition~\eqref{eq:chi3} for the cutoff
model like the PNJL model, $\chi_J$ is non-zero proportional to
$\Lambda^2$ even at $T=B=0$ as discussed in
Ref.~\cite{Fukushima:2009ft}.  This is in contradiction with the gauge
invariance, which requires the above susceptibility to vanish because
the current susceptibility is nothing but the $33$-component of the
photon polarization tensor at zero momentum.  Therefore, in order to
deal with the physically meaningful quantity, we subtract the vacuum
part from the above equation and compute,

\begin{equation}
 \bar\chi_J = \chi_J(\mu_5,B,T) - \chi_J(\mu_5,0,0) \;.
\label{eq:Cchi3}
\end{equation}
to fulfill the requirement that photons are unscreened at $T=B=0$
regardless of any value of $\mu_5$.

We plot our results for $\beta V \bar\chi_J$ as a function of $eB$
in Fig.~\ref{Fig:CurrentChi} at $\mu_5=0$ (upper panel),
$\mu_5=200\MeV$ (middle panel), and $\mu_5=400\MeV$ (lower panel).
The oscillations in the susceptibility behavior are an artifact of the
momentum cutoff.  In these plots $T_c = 228\MeV$ denotes the critical
temperature for chiral symmetry restoration at $\mu_5=B=0$.

Let us first focus on the case at $\mu_5 = 0$.  At $T=0$ and
$T=0.95T_c$ the system is in the broken phase with
$\langle\bar{u}u\rangle\neq0$ over the whole range of $eB$.  On
the other hand, at the temperature $T=1.1 T_c$, the system is in the
chiral symmetric phase for $eB$ smaller than a critical value.  There
is a phase transition from the symmetric to the broken phase with
increasing $eB$.  This transition is driven by the presence of the
magnetic field as the catalysis of chiral symmetry breaking, as
mentioned before.  The effect of the phase transition leads to a cusp
in the susceptibility $\chi_J$ as a function of $eB$.  We also notice
that there seems to exist a range in $eB$ in which $\bar\chi_J < 0$.
This is a mere artifact of the momentum cutoff, which causes
unphysical fluctuations in $\bar\chi_J$.  The qualitative picture is
similar also at $\mu_5\neq0$.

\begin{figure}
\begin{center}
\includegraphics[width=8cm]{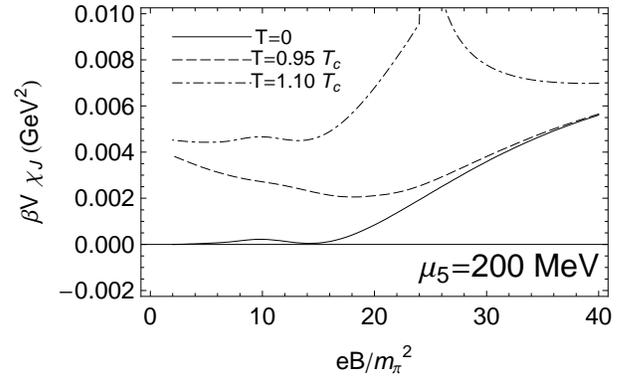}
\end{center}
\caption{\label{Fig:NewCurrentChi} Subtracted current susceptibility
  with a smoother regulator with $N=5$ in Eq.~\eqref{eq:f_p}.}
\end{figure}

 To distinguish physically meaningful information from
  cutoff artifacts, we have computed $\chi_J$ using a smoother
  regulator with $N=5$ in Eq.~\eqref{eq:f_p}.  We have readjusted the
  NJL parameters to keep the physical quantities ($f_\pi$ and
  $\langle\bar{u}u\rangle$) unchanged.  Figure~\ref{Fig:NewCurrentChi}
  is the result in which oscillations are suppressed and $\chi_J > 0$
  for any $T$ and $B$.

In view of Figs.~\ref{Fig:CurrentChi} and
  \ref{Fig:NewCurrentChi} we can conclude that it is certainly a
  physical effect that the chiral phase transition critically affects
  the susceptibility $\chi_J$ as well as $\chi_5$.  As shown in
  Ref.~\cite{Fukushima:2009ft} the susceptibility difference between
  the longitudinal and transverse directions has an origin in the
  axial anomaly and is insensitive to the infrared information.
  Nevertheless, $\chi_J$ (and transverse $\chi_J^T$ too) should be
  largely enhanced near the chiral phase transition through mixing
  with the divergent chiral susceptibility, which is not constrained
  by anomaly.  Such enhancement in $\chi_J$ would ease the
  confirmation of the CME signals at experiment.

\section{Conclusions}
\label{sec:conclusions}
In this article we have considered several aspects related to the
response of quark matter to an external magnetic field.  Quark matter
has been modelled by the Polyakov extended version of the
Nambu--Jona-Lasinio (PNJL) model, in which the QCD interaction among
quarks is replaced by effective four-fermion interactions.  In the
PNJL model, besides the quark-antiquark condensate which is
responsible for the dynamical chiral symmetry breaking in the QCD
vacuum, it is possible to compute the expectation value of the
Polyakov loop, which is a relevant indicator for the quark
deconfinement crossover.

In our study, we have firstly focused on the effect of a strong
magnetic field on chiral symmetry restoration at finite
temperature.  Our results show the effect of the external field as
a catalyzer of dynamical symmetry breaking.  Moreover, the critical
temperature increases as the strength of $B$ is increased.  This
behavior is in agreement with the previous studies on magnetic
catalysis in NJL-like models.

We have also discussed the effects of a chiral chemical potential,
$\mu_5$, on the phase structure of the model.  The chiral chemical
potential mimics the chirality induced by topological excitations
according to the QCD anomaly relation.  Instead of working at fixed
chirality $N_5$, we have worked in the grand-canonical ensemble
introducing $\mu_5$, i.e.\ the chemical potential conjugate to $N_5$.
Besides the phase diagram of the model, summarized in
Fig.~\ref{Fig:thePD}, we have computed several quantities that are
relevant for the Chiral Magnetic Effect (CME).  That is, we have
computed the current density $j_3$ and its susceptibility $\chi_J$ as
well as the chiral charge density $n_5$ and its susceptibility
$\chi_5$.

As a future project it is indispensable to extend our analysis to the
$2+1$ flavors and tune the PNJL model parameters to reproduce the
correct $T_c$ and thermodynamic properties, which would enable us to
make a serious comparison with the dynamical lattice-QCD
data~\cite{Abramczyk:2009gb}, and furthermore, it would be possible to
give more pertinent prediction on the physical observables.
\\

\acknowledgments M.~R.\ acknowledges discussions with H.~Abuki and
S.~Nicotri and K.~F.\ thanks E.~Fraga for discussions. The work of
M.~R.\ is supported by JSPS under the contract number P09028.
K.~F.\ is supported by Japanese MEXT grant No.\ 20740134 and also
supported in part by Yukawa International Program for Quark Hadron
Sciences.  The numerical calculations were carried out on Altix3700
BX2 at YITP in Kyoto University.



\begin{thebibliography}{99}

\bibitem{deForcrand:2006pv}
  P.~de Forcrand and O.~Philipsen,
  JHEP {\bf 0701}, 077 (2007)
  [arXiv:hep-lat/0607017];
  JHEP {\bf 0811}, 012 (2008)
  [arXiv:0808.1096 [hep-lat]];
  PoS {\bf LATTICE2008}, 208 (2008)
  [arXiv:0811.3858 [hep-lat]].

\bibitem{Aoki:2006br}
  Y.~Aoki, Z.~Fodor, S.~D.~Katz and K.~K.~Szabo,
  Phys.\ Lett.\  B {\bf 643}, 46 (2006)
  [arXiv:hep-lat/0609068];
  Y.~Aoki, S.~Borsanyi, S.~Durr, Z.~Fodor, S.~D.~Katz, S.~Krieg and K.~K.~Szabo,
  JHEP {\bf 0906}, 088 (2009)
  [arXiv:0903.4155 [hep-lat]].

\bibitem{Bazavov:2009zn}
  A.~Bazavov {\it et al.},
  Phys.\ Rev.\  D {\bf 80}, 014504 (2009)
  [arXiv:0903.4379 [hep-lat]].

\bibitem{Cheng:2009be}
  M.~Cheng {\it et al.},
  arXiv:0911.3450 [hep-lat].

\bibitem{Muroya:2003qs}
  S.~Muroya, A.~Nakamura, C.~Nonaka and T.~Takaishi,
  Prog.\ Theor.\ Phys.\  {\bf 110}, 615 (2003)
  [arXiv:hep-lat/0306031];
  K.~Splittorff,
  PoS {\bf LAT2006}, 023 (2006)
  [arXiv:hep-lat/0610072];
  K.~Splittorff and J.~J.~M.~Verbaarschot,
  Phys.\ Rev.\ Lett.\  {\bf 98}, 031601 (2007)
  [arXiv:hep-lat/0609076];
  K.~Fukushima and Y.~Hidaka,
  Phys.\ Rev.\  D {\bf 75}, 036002 (2007)
  [arXiv:hep-ph/0610323];
  J.~C.~R.~Bloch and T.~Wettig,
  JHEP {\bf 0903}, 100 (2009)
  [arXiv:0812.0324 [hep-lat]].

\bibitem{Fodor:2001pe}
  Z.~Fodor and S.~D.~Katz,
  JHEP {\bf 0203}, 014 (2002)
  [arXiv:hep-lat/0106002].

\bibitem{Allton:2005gk}
  C.~R.~Allton {\it et al.},
  Phys.\ Rev.\  D {\bf 71}, 054508 (2005)
  [arXiv:hep-lat/0501030];
  R.~V.~Gavai and S.~Gupta,
  Phys.\ Rev.\  D {\bf 78}, 114503 (2008)
  [arXiv:0806.2233 [hep-lat]].

\bibitem{Ambjorn:2002pz}
  J.~Ambjorn, K.~N.~Anagnostopoulos, J.~Nishimura and J.~J.~M.~Verbaarschot,
  JHEP {\bf 0210}, 062 (2002)
  [arXiv:hep-lat/0208025];
  Z.~Fodor, S.~D.~Katz and C.~Schmidt,
  JHEP {\bf 0703}, 121 (2007)
  [arXiv:hep-lat/0701022].

\bibitem{D'Elia:2002gd}
  M.~D'Elia and M.~P.~Lombardo,
  Phys.\ Rev.\  D {\bf 67}, 014505 (2003)
  [arXiv:hep-lat/0209146];
  Phys.\ Rev.\  D {\bf 70}, 074509 (2004)
  [arXiv:hep-lat/0406012];
  M.~D'Elia, F.~Di Renzo and M.~P.~Lombardo,
  Phys.\ Rev.\  D {\bf 76}, 114509 (2007)
  [arXiv:0705.3814 [hep-lat]];
  P.~de Forcrand and O.~Philipsen,
  Nucl.\ Phys.\  B {\bf 673}, 170 (2003)
  [arXiv:hep-lat/0307020].

\bibitem{Aarts:2008wh}
  G.~Aarts,
  Phys.\ Rev.\ Lett.\  {\bf 102}, 131601 (2009)
  [arXiv:0810.2089 [hep-lat]].

\bibitem{Kharzeev:2007jp}
  D.~E.~Kharzeev, L.~D.~McLerran and H.~J.~Warringa,
  Nucl.\ Phys.\  A {\bf 803}, 227 (2008)
  [arXiv:0711.0950 [hep-ph]].

\bibitem{Skokov:2009qp}
  V.~Skokov, A.~Y.~Illarionov and V.~Toneev,
  Int.\ J.\ Mod.\ Phys.\  A {\bf 24}, 5925 (2009)
  [arXiv:0907.1396 [nucl-th]].

\bibitem{Klevansky:1989vi}
  S.~P.~Klevansky and R.~H.~Lemmer,
  Phys.\ Rev.\  D {\bf 39}, 3478 (1989);
  H.~Suganuma and T.~Tatsumi,
  Annals Phys.\  {\bf 208}, 470 (1991);
  I.~A.~Shushpanov and A.~V.~Smilga,
  Phys.\ Lett.\  B {\bf 402}, 351 (1997)
  [arXiv:hep-ph/9703201];
  D.~N.~Kabat, K.~M.~Lee and E.~J.~Weinberg,
  Phys.\ Rev.\  D {\bf 66}, 014004 (2002)
  [arXiv:hep-ph/0204120];
  T.~Inagaki, D.~Kimura and T.~Murata,
  Prog.\ Theor.\ Phys.\  {\bf 111}, 371 (2004)
  [arXiv:hep-ph/0312005];
  T.~D.~Cohen, D.~A.~McGady and E.~S.~Werbos,
  Phys.\ Rev.\  C {\bf 76}, 055201 (2007)
  [arXiv:0706.3208 [hep-ph]];
  K.~Fukushima and H.~J.~Warringa,
  Phys.\ Rev.\ Lett.\  {\bf 100}, 032007 (2008)
  [arXiv:0707.3785 [hep-ph]];
  J.~L.~Noronha and I.~A.~Shovkovy,
  Phys.\ Rev.\  D {\bf 76}, 105030 (2007)
  [arXiv:0708.0307 [hep-ph]].

\bibitem{Gusynin:1995nb}
  V.~P.~Gusynin, V.~A.~Miransky and I.~A.~Shovkovy,
  Nucl.\ Phys.\  B {\bf 462}, 249 (1996)
  [arXiv:hep-ph/9509320];
  Nucl.\ Phys.\  B {\bf 563}, 361 (1999)
  [arXiv:hep-ph/9908320];
  G.~W.~Semenoff, I.~A.~Shovkovy and L.~C.~R.~Wijewardhana,
  Phys.\ Rev.\  D {\bf 60}, 105024 (1999)
  [arXiv:hep-th/9905116];
  V.~A.~Miransky and I.~A.~Shovkovy,
  Phys.\ Rev.\  D {\bf 66}, 045006 (2002)
  [arXiv:hep-ph/0205348].

\bibitem{Fraga:2008qn}
  E.~S.~Fraga and A.~J.~Mizher,
  Phys.\ Rev.\  D {\bf 78}, 025016 (2008)
  [arXiv:0804.1452 [hep-ph]];
  Nucl.\ Phys.\  A {\bf 831}, 91 (2009)
  [arXiv:0810.5162 [hep-ph]];
  J.~K.~Boomsma and D.~Boer,
  arXiv:0911.2164 [hep-ph].

\bibitem{Klimenko:1990rh}
  K.~G.~Klimenko,
  Theor.\ Math.\ Phys.\  {\bf 89}, 1161 (1992)
  [Teor.\ Mat.\ Fiz.\  {\bf 89}, 211 (1991)];
  K.~G.~Klimenko,
  Z.\ Phys.\  C {\bf 54}, 323 (1992);
  K.~G.~Klimenko,
  Theor.\ Math.\ Phys.\  {\bf 90}, 1 (1992)
  [Teor.\ Mat.\ Fiz.\  {\bf 90}, 3 (1992)].



\bibitem{Zayakin:2008cy}
  A.~V.~Zayakin,
  JHEP {\bf 0807}, 116 (2008)
  [arXiv:0807.2917 [hep-th]];
  G.~Lifschytz and M.~Lippert,
  Phys.\ Rev.\  D {\bf 80}, 066005 (2009)
  [arXiv:0904.4772 [hep-th]];
  Phys.\ Rev.\  D {\bf 80}, 066007 (2009)
  [arXiv:0906.3892 [hep-th]];
  H.~U.~Yee,
  JHEP {\bf 0911}, 085 (2009)
  [arXiv:0908.4189 [hep-th]];
  B.~Sahoo and H.~U.~Yee,
  arXiv:0910.5915 [hep-th];
  A.~Rebhan, A.~Schmitt and S.~A.~Stricker,
  arXiv:0909.4782 [hep-th];
  S.~l.~Cui, Y.~h.~Gao, Y.~Seo, S.~j.~Sin and W.~s.~Xu,
  arXiv:0910.2661 [hep-th];
  E.~D'Hoker and P.~Kraus,
  arXiv:0911.4518 [hep-th].

\bibitem{Gorbar:2005sx}
  E.~V.~Gorbar, S.~Homayouni and V.~A.~Miransky,
  Phys.\ Rev.\  D {\bf 72}, 065014 (2005)
  [arXiv:hep-th/0503028].

\bibitem{Fukushima:2008xe}
  K.~Fukushima, D.~E.~Kharzeev and H.~J.~Warringa,
  Phys.\ Rev.\  D {\bf 78}, 074033 (2008)
  [arXiv:0808.3382 [hep-ph]].

\bibitem{Arnold:1987zg}
  P.~Arnold and L.~D.~McLerran,
  Phys.\ Rev.\  D {\bf 37}, 1020 (1988).

\bibitem{Moore:1997im}
  G.~D.~Moore,
  Phys.\ Lett.\  B {\bf 412}, 359 (1997)
  [arXiv:hep-ph/9705248];
  arXiv:hep-ph/0009161;
  D.~Bodeker, G.~D.~Moore and K.~Rummukainen,
  Phys.\ Rev.\  D {\bf 61}, 056003 (2000)
  [arXiv:hep-ph/9907545].

\bibitem{McLerran:1990de}
  L.~D.~McLerran, E.~Mottola and M.~E.~Shaposhnikov,
  Phys.\ Rev.\  D {\bf 43}, 2027 (1991).

\bibitem{Voloshin:2004vk}
  S.~A.~Voloshin,
  Phys.\ Rev.\  C {\bf 70}, 057901 (2004)
  [arXiv:hep-ph/0406311].

\bibitem{STAR}
  B.~I.~Abelev {\it et al.}  [STAR Collaboration],
  arXiv:0909.1717 [nucl-ex];
  Phys.\ Rev.\ Lett.\  {\bf 103}, 251601 (2009)
  [arXiv:0909.1739 [nucl-ex]].

\bibitem{Nam:2009jb}
  S.~i.~Nam,
  Phys.\ Rev.\  D {\bf 80}, 114025 (2009)
  [arXiv:0911.0509 [hep-ph]].

\bibitem{Kharzeev:2009pj}
  D.~E.~Kharzeev and H.~J.~Warringa,
  Phys.\ Rev.\  D {\bf 80}, 034028 (2009)
  [arXiv:0907.5007 [hep-ph]].

\bibitem{Fukushima:2009ft}
  K.~Fukushima, D.~E.~Kharzeev and H.~J.~Warringa,
  arXiv:0912.2961 [hep-ph].

\bibitem{Buividovich:2009wi}
  P.~V.~Buividovich, M.~N.~Chernodub, E.~V.~Luschevskaya and M.~I.~Polikarpov,
  Phys.\ Rev.\  D {\bf 80}, 054503 (2009)
  [arXiv:0907.0494 [hep-lat]].

\bibitem{Abramczyk:2009gb}
  M.~Abramczyk, T.~Blum, G.~Petropoulos and R.~Zhou,
  arXiv:0911.1348 [hep-lat].

\bibitem{Meisinger:1995ih}
  P.~N.~Meisinger and M.~C.~Ogilvie,
  Phys.\ Lett.\  B {\bf 379}, 163 (1996)
  [arXiv:hep-lat/9512011].

\bibitem{Fukushima:2003fw}
  K.~Fukushima,
  Phys.\ Lett.\  B {\bf 591}, 277 (2004)
  [arXiv:hep-ph/0310121].

\bibitem{revNJL}
  U.~Vogl and W.~Weise,
  Prog.\ Part.\ Nucl.\ Phys.\  {\bf 27}, 195 (1991);
  S.~P.~Klevansky,
  Rev.\ Mod.\ Phys.\  {\bf 64}, 649 (1992);
  T.~Hatsuda and T.~Kunihiro,
  Phys.\ Rept.\  {\bf 247}, 221 (1994)
  [arXiv:hep-ph/9401310];
  M.~Buballa,
  Phys.\ Rept.\  {\bf 407}, 205 (2005)
  [arXiv:hep-ph/0402234].

\bibitem{Polyakovetal}
  A.~M.~Polyakov,
  Phys.\ Lett.\  B {\bf 72}, 477 (1978);
  L.~Susskind,
  Phys.\ Rev.\  D {\bf 20}, 2610 (1979);
  B.~Svetitsky and L.~G.~Yaffe,
  Nucl.\ Phys.\  B {\bf 210}, 423 (1982);
  B.~Svetitsky,
  Phys.\ Rept.\  {\bf 132}, 1 (1986).

\bibitem{Ratti:2005jh}
  C.~Ratti, M.~A.~Thaler and W.~Weise,
  Phys.\ Rev.\  D {\bf 73}, 014019 (2006)
  [arXiv:hep-ph/0506234];
  E.~Megias, E.~Ruiz Arriola and L.~L.~Salcedo,
  Phys.\ Rev.\  D {\bf 74}, 114014 (2006)
  [arXiv:hep-ph/0607338];
  Eur.\ Phys.\ J.\  A {\bf 31}, 553 (2007)
  [arXiv:hep-ph/0610163];
  S.~Roessner, C.~Ratti and W.~Weise,
  Phys.\ Rev.\  D {\bf 75}, 034007 (2007)
  [arXiv:hep-ph/0609281];
  C.~Sasaki, B.~Friman and K.~Redlich,
  Phys.\ Rev.\  D {\bf 75}, 074013 (2007)
  [arXiv:hep-ph/0611147];
  S.~K.~Ghosh, T.~K.~Mukherjee, M.~G.~Mustafa and R.~Ray,
  Phys.\ Rev.\  D {\bf 77}, 094024 (2008)
  [arXiv:0710.2790 [hep-ph]];
  W.~j.~Fu, Z.~Zhang and Y.~x.~Liu,
  Phys.\ Rev.\  D {\bf 77}, 014006 (2008)
  [arXiv:0711.0154 [hep-ph]];
  M.~Ciminale, R.~Gatto, N.~D.~Ippolito, G.~Nardulli and M.~Ruggieri,
  Phys.\ Rev.\  D {\bf 77}, 054023 (2008)
  [arXiv:0711.3397 [hep-ph]];
  Y.~Sakai, K.~Kashiwa, H.~Kouno and M.~Yahiro,
  Phys.\ Rev.\  D {\bf 77}, 051901 (2008)
  [arXiv:0801.0034 [hep-ph]];
  Phys.\ Rev.\  D {\bf 78}, 036001 (2008)
  [arXiv:0803.1902 [hep-ph]];
  K.~Kashiwa, H.~Kouno and M.~Yahiro,
  arXiv:0908.1213 [hep-ph];
  K.~Fukushima,
  Phys.\ Rev.\  D {\bf 77}, 114028 (2008)
  [Erratum-ibid.\  D {\bf 78}, 039902 (2008)]
  [arXiv:0803.3318 [hep-ph]];
  H.~Abuki, R.~Anglani, R.~Gatto, G.~Nardulli and M.~Ruggieri,
  Phys.\ Rev.\  D {\bf 78}, 034034 (2008)
  [arXiv:0805.1509 [hep-ph]];
  H.~Abuki, M.~Ciminale, R.~Gatto and M.~Ruggieri,
  Phys.\ Rev.\  D {\bf 79}, 034021 (2009)
  [arXiv:0811.1512 [hep-ph]];
  T.~Hell, S.~Roessner, M.~Cristoforetti and W.~Weise,
  Phys.\ Rev.\  D {\bf 79}, 014022 (2009)
  [arXiv:0810.1099 [hep-ph]];
  arXiv:0911.3510 [hep-ph].

\bibitem{Schaefer:2007pw}
  B.~J.~Schaefer, J.~M.~Pawlowski and J.~Wambach,
  Phys.\ Rev.\  D {\bf 76}, 074023 (2007)
  [arXiv:0704.3234 [hep-ph]].
\bibitem{Braun:2009gm}
  J.~Braun, L.~M.~Haas, F.~Marhauser and J.~M.~Pawlowski,
  arXiv:0908.0008 [hep-ph].

\bibitem{Agasian:2008tb}
  N.~O.~Agasian and S.~M.~Fedorov,
  Phys.\ Lett.\  B {\bf 663}, 445 (2008)
  [arXiv:0803.3156 [hep-ph]].

\bibitem{Fukushima:2010vw}
  K.~Fukushima, D.~E.~Kharzeev and H.~J.~Warringa,
  arXiv:1002.2495 [hep-ph].

\bibitem{Wambach:2009ee}
  J.~Wambach, B.~J.~Schaefer and M.~Wagner,
  arXiv:0911.0296 [hep-ph].

\bibitem{Ritus:1972ky}
  V.~I.~Ritus,
  Annals Phys.\  {\bf 69}, 555 (1972);
  Sov.\ Phys.\ JETP {\bf 48}, 788 (1978)
  [Zh.\ Eksp.\ Teor.\ Fiz.\  {\bf 75}, 1560 (1978)].

\bibitem{Campanelli:2009sc}
  L.~Campanelli and M.~Ruggieri,
  Phys.\ Rev.\  D {\bf 80}, 034014 (2009)
  [arXiv:0905.0853 [hep-ph]].

\end{thebibliography}
\end{document}